\documentclass[aps,twocolumn,showpacs,preprintnumbers,floats]{revtex4}\def\@cite#1#2{\textsuperscript{[{#1\if@tempswa , #2\fi}]}}

\usepackage{graphicx}
\usepackage{amsmath}
\usepackage{amsfonts}
\usepackage{amssymb}
\usepackage{color}
\usepackage{subfigure}
\usepackage{epsfig}
\usepackage{morefloats}
\usepackage{multirow}
\usepackage{graphicx,booktabs}
\usepackage{mathrsfs}
\usepackage{txfonts}
\usepackage{indentfirst}
\usepackage{graphicx,booktabs}

\usepackage{longtable,lscape}

\newcommand{\vsig}{\mbox{\boldmath$\sigma$\unboldmath}}
\newcommand{\veps}{\mbox{\boldmath$\epsilon$\unboldmath}}

\begin{document}

\title{\textbf{ The excited bottom-charmed mesons in a nonrelativistic quark model}}
\author{
Qi Li, Ming-Sheng Liu, Long-Sheng Lu, Qi-Fang L\"{u} \footnote {E-mail: lvqifang@hunnu.edu.cn}, Long-Cheng Gui~\footnote{E-mail: guilongcheng@hunnu.edu.cn}, and Xian-Hui Zhong
\footnote {E-mail: zhongxh@hunnu.edu.cn} }  \affiliation{ 1) Department
of Physics, Hunan Normal University,  Changsha 410081, China }

\affiliation{ 2) Synergetic Innovation
Center for Quantum Effects and Applications (SICQEA), Changsha 410081,China}

\affiliation{ 3) Key Laboratory of
Low-Dimensional Quantum Structures and Quantum Control of Ministry
of Education, Changsha 410081, China}


\begin{abstract}

Using the newly measured masses of $B_c(1S)$ and $B_c(2S)$ from the CMS Collaboration and the $1S$ hyperfine splitting determined from the lattice QCD as constrains, we calculate the $B_c$ mass spectrum up to the $6S$ multiplet with a nonrelativistic linear potential model.
Furthermore, using the wave functions from this model we calculate the radiative transitions between the $B_c$ states within a constituent quark model. For the higher mass $B_c$ states lying above $DB$ threshold, we also evaluate the Okubo-Zweig-Iizuka (OZI)  allowed two-body strong decays with the $^{3}P_{0}$ model. Our study indicates that besides there are large potentials for the observations of the low-lying $B_c$ states below the $DB$ threshold via their radiative transitions, some higher mass $B_c$ states, such as $B_c(2^3P_2)$, $B_c(2^3D_1)$, $B_c(3^3D_1)$, $B_c(4^3P_0)$, and the $1F$-wave $B_c$ states, might be first observed in their dominant strong decay channels $DB$, $DB^*$ or $D^*B$ at the LHC for their relatively narrow widths.

\end{abstract}

\maketitle

\section{Introduction}

The $B_c$ states composed of a bottom-charmed quark-antiquark pair, as an important family of hadron spectra was predicted in theory about 40 years ago~\cite{Eichten:1980mw}, however,
the experimental progress towards establishing the $B_c$ spectrum is not obvious. Except for the ground state $B_c$ meson observed in 1998 by the CDF Collaboration at Fermilab~\cite{Abe:1998fb}, until 2018, only the ATLAS Collaboration reported evidence of
an excited $B_c$ state with a mass of $6842\pm 9$ MeV~\cite{Aad:2014laa} consistent with the values predicted for $B_c(2S)$, while it was not confirmed by the LHCb Collaboration by using their 8 TeV data sample~\cite{Aaij:2017lpg}. The poor situation of the
observations and measurements of $B_c$ spectrum is due to that the production yields are significantly smaller than those of the charmonium and bottomonium ($c\bar{c}$ and $b\bar{b}$) states. Fortunately, the LHC provides good opportunities for our search for the excited $B_c$ states with its high collision energies and integrated luminosity. Very recently, two excited $B_c^+$ states were observed in the $B_c^+\pi^+\pi^-$ invariant
mass spectrum by the CMS Collaboration~\cite{Sirunyan:2019osb}. Signals are consistent with the $B_c(2S)$ and
$B^*_c(2S)$ states. These two states are well resolved from each other and are observed with a significance
exceeding five standard deviations. The mass of $B_c(2S)$ meson, $6871\pm 2.8$ MeV, measured by the CMS
Collaboration is inconsistent with the determination $6842\pm 9$ MeV by the ATLAS Collaboration. The reason is that the peak observed
by ATLAS could be the superposition of the $B_c(2S)$ and $B^*_c(2S)$ states, too closely spaced with
respect to the resolution of the measurement~\cite{Sirunyan:2019osb}.

The $B_c$ states as the only conventional heavy mesons with different
flavors have aroused great interests in theory. Compared with the $c\bar{c}$ and $b\bar{b}$ spectra, the $B_c$ spectrum has several special features for the bottom-charmed quark-antiquark pair. (i) The $B_c$ states cannot annihilate into gluons, thus, the lowlying excited $B_c$ states below the $DB$ threshold are more stable with a narrow width less than a few hundred keV, they mainly decay via the electromagnetic or hadronic transitions between two different $B_c$ states. (ii) In the $B_c$ meson spectrum there are configuration mixings between the states with different total spins but with the same total angular momentum, such as $^3P_1-^1P_1$, $^3D_2-^1D_2$, and $^3F_3-^1F_3$ mixings via the antisymmetric part of the spin-orbit potential.
(iii) Additionally, the $B_c$ states provide a unique window for studying the heavy-quark dynamics that is very different from those provided by the $c\bar{c}$ and $b\bar{b}$ states. In the past years, the $B_c$ mass spectrum has been predicted with various models~\cite{Zeng:1994vj,Eichten:1994gt,Ebert:2002pp,
Godfrey:2004ya,Kiselev:1994rc,Monteiro:2016ijw,Soni:2017wvy,Godfrey:1985xj,Tang:2018myz,Baldicchi:2000cf,Ikhdair:2003tt,
Ikhdair:2003ry,Wang:2012kw,Chen:2011qu,Wei:2010zza,Badalian:2009cx,Rai:2008sc,Guo:2008he,Patel:2008na,Bernotas:2008bu,
AbdElHady:2005bv,AbdElHady:1998kc,Li:2004gu,Ikhdair:2004hg,Fulcher:1998ka,Motyka:1997di,Devlani:2014nda,Monteiro:2016rzi,Eichten:2019gig}. Furthermore, a few lattice calculations can be found in Refs.~\cite{Davies:1996gi,Mathur:2018epb,Gregory:2009hq,Dowdall:2012ab,Allison:2004be}. To estimate the production rates in experiments, the production of the excited $B_c$ states were often discussed in the literature~\cite{Braaten:1996pv,Cheung:1995ir,Cheung:1995ye,Cheung:1993qi,Berezhnoy:1996ks,Chang:2004bh,Chang:2005bf,Liao:2018nab,Kai:2017cba,Liao:2015vqa,Liao:2014rca,Liao:2011kd,Yang:2010yg,Chang:2007si}. As the dominant decay modes, the electromagnetic transitions of the low-lying $B_c$ states were also widely estimated in the literature~\cite{Eichten:1994gt,Zeng:1994vj,Kiselev:1994rc,Fulcher:1998ka,Ebert:2002pp,Godfrey:2004ya,
AbdElHady:2005bv,Devlani:2014nda,Monteiro:2016ijw,Soni:2017wvy,Patnaik:2017cbl,Simonis:2016pnh,Wang:2015yea,Wang:2013cha,Ebert:2002xz}. However, the studies of the OZI-allowed strong decays for the high-lying $B_c$ states are confined only to a few calculations~\cite{Kiselev:1996un,Ferretti:2015rsa,Monteiro:2016rzi,Eichten:2019gig}.

The successes of the observations of the radially excited $B_c$ states $B_c(2S)$ and $B^*_c(2S)$ by the CMS Collaboration~\cite{Sirunyan:2019osb} have demonstrated that more excited $B_c$ states are to be discovered in future LHC experiments. Stimulated by the great discovery potentials of the missing $B_c$ states in future experiments, in present work we carry out a systematical study of the $B_c$ spectrum. First, using the newly measured masses of $B_c(1S)$ and $B_c(2S)$ from the CMS Collaboration~\cite{Sirunyan:2019osb} and the $1S$ hyperfine splitting determined from the lattice QCD~\cite{Gregory:2009hq,Dowdall:2012ab,Mathur:2018epb} as constrains, we calculate the $B_c$ mass spectrum up to the $6S$ multiplet with a nonrelativistic linear potential model. The slope parameter of the linear potential has been well determined in our previous study of the charmonium states~\cite{Deng:2016stx}.  To involve the spin-dependent corrections of the spatial wave functions, following the method adopted in Refs.~\cite{Deng:2016ktl,Deng:2016stx} we treat the spin-dependent interactions as nonperturbative terms in our calculations. With this nonperturbative treatment, we can reasonably include the effect of spin-dependent interactions on the spatial wave functions, which is essential for us to gain reliable predictions of the decay behaviors.

Then, with the available wavefunctions from the potential model, we evaluate the electromagnetic transitions between the $B_c$ states within a nonrelativistic constituent quark model developed in our previous works~\cite{Deng:2016ktl,Deng:2016stx}. With this approach the possible higher EM multipole contributions to a EM transition process can be included naturally. Considering the fact that the higher $B_c$ states lying above the $DB$ threshold may have enough possibilities to be produced at LHC, and they are easy to be established in the $D^{(*)}B^{(*)}$ hadronic final states, thus to give useful references for the LHC observations, we further calculate the OZI-allowed strong decays of the higher $B_c$ states within the widely used $^3P_0$ model~\cite{Micu:1968mk,LeYaouanc:1972vsx,LeYaouanc:1973ldf}. It is found that $B_c(2^3P_2)$, $B_c(2^3D_1)$, $B_c(3^3D_1)$ together with the $1F$-wave $B_c$ states might be first observed in their dominant strong decay channels $DB$, $DB^*$ or $D^*B$ at LHC for their relatively narrow width.

This paper is organized as follows. In Sec.~\ref{spectrum}, the $B_c$ mass spectrum is calculated within a nonrelativistic linear potential model.
Then, with the obtained $B_c$ spectrum the radiative transitions between the $B_c$ states are estimated in Sec.~\ref{EMT} within a nonrelativistic constituent quark model. In Sec.~\ref{Strongdecay}, the OZI-allowed two-body strong decays of the excited $B_c$ state are also studied within the $^3P_0$ model. In Sec.~\ref{DIS}, we focus on the calculation results and discuss some strategies for looking for the $B_c$ states in future experiments. Finally, a summary is given in Sec.~\ref{sum}.

\begin{table*} 
\caption{Predicted masses (MeV) of $B_c$ states compared with other model predictions and data.
The mixing angles between $B_{c}(n^{3}L_{J})$ and $B_{c}(n^{1}L_{J})$ obtained in this work are presented in Table~\ref{mixangle}.}\label{mass}
\begin{tabular}{ccccccccccccccccc}\midrule[1.0pt]\midrule[1.0pt]
& State~~~~&$J^{P}$~~~~ &Ours~~~~ & ZVR~\cite{Zeng:1994vj}~~~~& SJSCP~\cite{Soni:2017wvy} &MBV~\cite{Monteiro:2016ijw}~~~~&EQ~\cite{Eichten:1994gt}~~~~
&EFG~\cite{Ebert:2002pp}~~~~& GI~\cite{Godfrey:2004ya}~~~~& KLT~\cite{Kiselev:1994rc}~~~~& Lattice~\cite{Mathur:2018epb}&Exp~\cite{Sirunyan:2019osb}~~~~\\
\midrule[1.0pt]\										
&$B_{c}(1 ^3S_{1})$~~~~&$1^{-}$~~~~&6326 (input)~~~~~&6340    ~~~~&6321    ~~~~ &6357     ~~~~&6337    ~~~~&6332    ~~~~&6338    ~~~~&6317    ~~~~&6331$\pm 10$~~~~ &$\cdots$~~~~\\
&$B_{c}(1 ^1S_{0})$~~~~&$0^{-}$~~~~&6271 (input)~~~~~&6260    ~~~~&6272    ~~~~ &6275     ~~~~&6264    ~~~~&6270    ~~~~&6271    ~~~~&6253    ~~~~&6276$\pm9~$~~~~  &6271    ~~~~\\
&$B_{c}(2 ^3S_{1})$~~~~&$1^{-}$~~~~&6890        ~~~~~&6900    ~~~~&6900    ~~~~ &6897     ~~~~&6899    ~~~~&6881    ~~~~&6887    ~~~~&6902    ~~~~&$\cdots$~~~~     &$\cdots$~~~~\\
&$B_{c}(2 ^1S_{0})$~~~~&$0^{-}$~~~~&6871 (input)~~~~~&6850    ~~~~&6864    ~~~~ &6862     ~~~~&6856    ~~~~&6835    ~~~~&6855    ~~~~&6867    ~~~~&        ~~~~     &6871    ~~~~\\
&$B_{c}(3 ^3S_{1})$~~~~&$1^{-}$~~~~&7252        ~~~~~&7280    ~~~~&7338    ~~~~ &7333     ~~~~&7280    ~~~~&7235    ~~~~&7272    ~~~~&$\cdots$~~~~&$\cdots$~~~~     &$\cdots$~~~~\\
&$B_{c}(3 ^1S_{0})$~~~~&$0^{-}$~~~~&7239        ~~~~~&7240    ~~~~&7306    ~~~~ &7308     ~~~~&7244    ~~~~&7193    ~~~~&7250    ~~~~&$\cdots$~~~~&$\cdots$~~~~     &$\cdots$~~~~\\
&$B_{c}(4 ^3S_{1})$~~~~&$1^{-}$~~~~&7550        ~~~~~&7580    ~~~~&7714    ~~~~ &7734     ~~~~&7594    ~~~~&$\cdots$~~~~&$\cdots$~~~~&$\cdots$~~~~&$\cdots$~~~~     &$\cdots$~~~~\\
&$B_{c}(4 ^1S_{0})$~~~~&$0^{-}$~~~~&7540        ~~~~~&7550    ~~~~&7684    ~~~~ &7713     ~~~~&7562    ~~~~&$\cdots$~~~~&$\cdots$~~~~&$\cdots$~~~~&$\cdots$~~~~     &$\cdots$~~~~\\
&$B_{c}(5 ^3S_{1})$~~~~&$1^{-}$~~~~&7813        ~~~~~&$\cdots$~~~~&8054    ~~~~ &8115     ~~~~&$\cdots$~~~~&$\cdots$~~~~&$\cdots$~~~~&$\cdots$~~~~&$\cdots$~~~~     &$\cdots$~~~~\\
&$B_{c}(5 ^1S_{0})$~~~~&$0^{-}$~~~~&7805        ~~~~~&$\cdots$~~~~&8025    ~~~~ &8097     ~~~~&$\cdots$~~~~&$\cdots$~~~~&$\cdots$~~~~&$\cdots$~~~~&$\cdots$~~~~     &$\cdots$~~~~\\
&$B_{c}(6 ^3S_{1})$~~~~&$1^{-}$~~~~&8054        ~~~~~&$\cdots$~~~~&8368    ~~~~ &8484     ~~~~&$\cdots$~~~~&$\cdots$~~~~&$\cdots$~~~~&$\cdots$~~~~&$\cdots$~~~~     &$\cdots$~~~~\\
&$B_{c}(6 ^1S_{0})$~~~~&$0^{-}$~~~~&8046        ~~~~~&$\cdots$~~~~&8340    ~~~~ &8469     ~~~~&$\cdots$~~~~&$\cdots$~~~~&$\cdots$~~~~&$\cdots$~~~~&$\cdots$~~~~     &$\cdots$~~~~\\
&$B_{c}(1 ^3P_{2})$~~~~&$2^{+}$~~~~&6787        ~~~~~&6760    ~~~~&6712    ~~~~ &6737     ~~~~&6747    ~~~~&6762    ~~~~&6768    ~~~~&6743    ~~~~&$\cdots$~~~~     &$\cdots$~~~~\\
&$B_{c}(1P'_1)$    ~~~~&$1^{+}$~~~~&6776        ~~~~~&6740    ~~~~&$\cdots$~~~~ &6734     ~~~~&6736    ~~~~&6749    ~~~~&6750    ~~~~&6729    ~~~~&$\cdots$~~~~     &$\cdots$~~~~\\
&$B_{c}(1P_1)$     ~~~~&$1^{+}$~~~~&6757        ~~~~~&6730    ~~~~&$\cdots$~~~~ &6686     ~~~~&6730    ~~~~&6734    ~~~~&6741    ~~~~&6717    ~~~~&$6736\pm 24$~~~~ &$\cdots$~~~~\\
&$B_{c}(1 ^3P_{0})$~~~~&$0^{+}$~~~~&6714        ~~~~~&6680    ~~~~&6686    ~~~~ &6638     ~~~~&6700    ~~~~&6699    ~~~~&6706    ~~~~&6683    ~~~~&$6712\pm25$~~~~  &$\cdots$~~~~\\
&$B_{c}(2 ^3P_{2})$~~~~&$2^{+}$~~~~&7160        ~~~~~&7160    ~~~~&7173    ~~~~ &7175     ~~~~&7153    ~~~~&7156    ~~~~&7164    ~~~~&7134    ~~~~&$\cdots$~~~~     &$\cdots$~~~~\\
&$B_{c}(2P'_1)$    ~~~~&$1^{+}$~~~~&7150        ~~~~~&7150    ~~~~&$\cdots$~~~~ &7173     ~~~~&7142    ~~~~&7145    ~~~~&7150    ~~~~&7124    ~~~~&$\cdots$~~~~     &$\cdots$~~~~\\
&$B_{c}(2P_1)$     ~~~~&$1^{+}$~~~~&7134        ~~~~~&7140    ~~~~&$\cdots$~~~~ &7137     ~~~~&7135    ~~~~&7126    ~~~~&7145    ~~~~&7113    ~~~~&$\cdots$~~~~     &$\cdots$~~~~\\
&$B_{c}(2 ^3P_{0})$~~~~&$0^{+}$~~~~&7107        ~~~~~&7100    ~~~~&7146    ~~~~ &7084     ~~~~&7108    ~~~~&7091    ~~~~&7122    ~~~~&7088    ~~~~&$\cdots$~~~~     &$\cdots$~~~~\\
&$B_{c}(3 ^3P_{2})$~~~~&$2^{+}$~~~~&7464        ~~~~~&7480    ~~~~&7565    ~~~~ &7575     ~~~~&$\cdots$~~~~&$\cdots$~~~~&$\cdots$~~~~&$\cdots$~~~~&$\cdots$~~~~     &$\cdots$~~~~\\
&$B_{c}(3P'_1)$    ~~~~&$1^{+}$~~~~&7458        ~~~~~&7470    ~~~~&$\cdots$~~~~ &7572     ~~~~&$\cdots$~~~~&$\cdots$~~~~&$\cdots$~~~~&$\cdots$~~~~&$\cdots$~~~~     &$\cdots$~~~~\\
&$B_{c}(3P_1)$     ~~~~&$1^{+}$~~~~&7441        ~~~~~&7460    ~~~~&$\cdots$~~~~ &7546     ~~~~&$\cdots$~~~~&$\cdots$~~~~&$\cdots$~~~~&$\cdots$~~~~&$\cdots$~~~~     &$\cdots$~~~~\\
&$B_{c}(3 ^3P_{0})$~~~~&$0^{+}$~~~~&7420        ~~~~~&7430    ~~~~&7536    ~~~~ &7492     ~~~~&$\cdots$~~~~&$\cdots$~~~~&$\cdots$~~~~&$\cdots$~~~~&$\cdots$~~~~     &$\cdots$~~~~\\
&$B_{c}(4 ^3P_{2})$~~~~&$2^{+}$~~~~&7732        ~~~~~&7760    ~~~~&7915    ~~~~ &7970     ~~~~&$\cdots$~~~~&$\cdots$~~~~&$\cdots$~~~~&$\cdots$~~~~&$\cdots$~~~~     &$\cdots$~~~~\\
&$B_{c}(4P'_1)$    ~~~~&$1^{+}$~~~~&7727        ~~~~~&7740    ~~~~&$\cdots$~~~~ &7942     ~~~~&$\cdots$~~~~&$\cdots$~~~~&$\cdots$~~~~&$\cdots$~~~~&$\cdots$~~~~     &$\cdots$~~~~\\
&$B_{c}(4P_1)$     ~~~~&$1^{+}$~~~~&7710        ~~~~~&7740    ~~~~&$\cdots$~~~~ &7943     ~~~~&$\cdots$~~~~&$\cdots$~~~~&$\cdots$~~~~&$\cdots$~~~~&$\cdots$~~~~     &$\cdots$~~~~\\
&$B_{c}(4 ^3P_{0})$~~~~&$0^{+}$~~~~&7693        ~~~~~&7710    ~~~~&7885    ~~~~ &7970     ~~~~&$\cdots$~~~~&$\cdots$~~~~&$\cdots$~~~~&$\cdots$~~~~&$\cdots$~~~~     &$\cdots$~~~~\\
&$B_{c}(1 ^3D_{3})$~~~~&$3^{-}$~~~~&7030        ~~~~~&7040    ~~~~&6990    ~~~~ &7004     ~~~~&7005    ~~~~&7081    ~~~~&7045    ~~~~&7134    ~~~~&$\cdots$~~~~     &$\cdots$~~~~\\
&$B_{c}(1D'_2)$    ~~~~&$2^{-}$~~~~&7032        ~~~~~&7030    ~~~~&$\cdots$~~~~ &7003     ~~~~&7012    ~~~~&7079    ~~~~&7036    ~~~~&7124    ~~~~&$\cdots$~~~~     &$\cdots$~~~~\\
&$B_{c}(1D_2)$     ~~~~&$2^{-}$~~~~&7024        ~~~~~&7020    ~~~~&$\cdots$~~~~ &6974     ~~~~&7009    ~~~~&7077    ~~~~&7041    ~~~~&7113    ~~~~&$\cdots$~~~~     &$\cdots$~~~~\\
&$B_{c}(1 ^3D_{1})$~~~~&$1^{-}$~~~~&7020        ~~~~~&7010    ~~~~&6998    ~~~~ &6973     ~~~~&7012    ~~~~&7072    ~~~~&7025    ~~~~&7088    ~~~~&$\cdots$~~~~     &$\cdots$~~~~\\
&$B_{c}(2 ^3D_{3})$~~~~&$3^{-}$~~~~&7348        ~~~~~&7370    ~~~~&7399    ~~~~ &7410     ~~~~&$\cdots$~~~~&$\cdots$~~~~&$\cdots$~~~~&$\cdots$~~~~&$\cdots$~~~~     &$\cdots$~~~~\\
&$B_{c}(2D'_2)$    ~~~~&$2^{-}$~~~~&7347        ~~~~~&7360    ~~~~&$\cdots$~~~~ &7408     ~~~~&$\cdots$~~~~&$\cdots$~~~~&$\cdots$~~~~&$\cdots$~~~~&$\cdots$~~~~     &$\cdots$~~~~\\
&$B_{c}(2D_2)$     ~~~~&$2^{-}$~~~~&7343        ~~~~~&7360    ~~~~&$\cdots$~~~~ &7385     ~~~~&$\cdots$~~~~&$\cdots$~~~~&$\cdots$~~~~&$\cdots$~~~~&$\cdots$~~~~     &$\cdots$~~~~\\
&$B_{c}(2 ^3D_{1})$~~~~&$1^{-}$~~~~&7336        ~~~~~&7350    ~~~~&7403    ~~~~ &7377     ~~~~&$\cdots$~~~~&$\cdots$~~~~&$\cdots$~~~~&$\cdots$~~~~&$\cdots$~~~~     &$\cdots$~~~~\\
&$B_{c}(3 ^3D_{3})$~~~~&$3^{-}$~~~~&7625        ~~~~~&7660    ~~~~&7761    ~~~~ &7796     ~~~~&$\cdots$~~~~&$\cdots$~~~~&$\cdots$~~~~&$\cdots$~~~~&$\cdots$~~~~     &$\cdots$~~~~\\
&$B_{c}(3D'_2)$    ~~~~&$2^{-}$~~~~&7623        ~~~~~&7650    ~~~~&$\cdots$~~~~ &7783     ~~~~&$\cdots$~~~~&$\cdots$~~~~&$\cdots$~~~~&$\cdots$~~~~&$\cdots$~~~~     &$\cdots$~~~~\\
&$B_{c}(3D_2)$     ~~~~&$2^{-}$~~~~&7620        ~~~~~&7650    ~~~~&$\cdots$~~~~ &7781     ~~~~&$\cdots$~~~~&$\cdots$~~~~&$\cdots$~~~~&$\cdots$~~~~&$\cdots$~~~~     &$\cdots$~~~~\\
&$B_{c}(3 ^3D_{1})$~~~~&$1^{-}$~~~~&7611        ~~~~~&7640    ~~~~&7762    ~~~~ &7761     ~~~~&$\cdots$~~~~&$\cdots$~~~~&$\cdots$~~~~&$\cdots$~~~~&$\cdots$~~~~     &$\cdots$~~~~\\
&$B_{c}(1 ^3F_{4})$~~~~&$4^{+}$~~~~&7227        ~~~~~&7250    ~~~~&7244    ~~~~ &$\cdots$ ~~~~&$\cdots$~~~~&$\cdots$~~~~&7271    ~~~~&$\cdots$~~~~&$\cdots$~~~~     &$\cdots$~~~~\\
&$B_{c}(1F'_3)$    ~~~~&$3^{+}$~~~~&7240        ~~~~~&7250    ~~~~&$\cdots$~~~~ &$\cdots$ ~~~~&$\cdots$~~~~&$\cdots$~~~~&7266    ~~~~&$\cdots$~~~~&$\cdots$~~~~     &$\cdots$~~~~\\
&$B_{c}(1F_3)$     ~~~~&$3^{+}$~~~~&7224        ~~~~~&7240    ~~~~&$\cdots$~~~~ &$\cdots$ ~~~~&$\cdots$~~~~&$\cdots$~~~~&7276    ~~~~&$\cdots$~~~~&$\cdots$~~~~     &$\cdots$~~~~\\
&$B_{c}(1 ^3F_{2})$~~~~&$2^{+}$~~~~&7235        ~~~~~&7240    ~~~~&7234    ~~~~ &$\cdots$ ~~~~&$\cdots$~~~~&$\cdots$~~~~&7269    ~~~~&$\cdots$~~~~&$\cdots$~~~~     &$\cdots$~~~~\\
&$B_{c}(2 ^3F_{4})$~~~~&$4^{+}$~~~~&7514        ~~~~~&7550    ~~~~&7617    ~~~~ &$\cdots$ ~~~~&$\cdots$~~~~&$\cdots$~~~~&7568    ~~~~&$\cdots$~~~~&$\cdots$~~~~     &$\cdots$~~~~\\
&$B_{c}(2F'_3)$    ~~~~&$3^{+}$~~~~&7525        ~~~~~&7550    ~~~~&$\cdots$~~~~ &$\cdots$ ~~~~&$\cdots$~~~~&$\cdots$~~~~&7571    ~~~~&$\cdots$~~~~&$\cdots$~~~~     &$\cdots$~~~~\\
&$B_{c}(2F_3)$     ~~~~&$3^{+}$~~~~&7508        ~~~~~&7540    ~~~~&$\cdots$~~~~ &$\cdots$ ~~~~&$\cdots$~~~~&$\cdots$~~~~&7563    ~~~~&$\cdots$~~~~&$\cdots$~~~~     &$\cdots$~~~~\\
&$B_{c}(2 ^3F_{2})$~~~~&$2^{+}$~~~~&7518        ~~~~~&7540    ~~~~&7607    ~~~~ &$\cdots$ ~~~~&$\cdots$~~~~&$\cdots$~~~~&7565    ~~~~&$\cdots$~~~~&$\cdots$~~~~     &$\cdots$~~~~\\
&$B_{c}(3 ^3F_{4})$~~~~&$4^{+}$~~~~&7771        ~~~~~&7810    ~~~~&7956    ~~~~ &$\cdots$ ~~~~&$\cdots$~~~~&$\cdots$~~~~&$\cdots$~~~~&$\cdots$~~~~&$\cdots$~~~~     &$\cdots$~~~~\\
&$B_{c}(3F'_3)$    ~~~~&$3^{+}$~~~~&7779        ~~~~~&7810    ~~~~&$\cdots$~~~~ &$\cdots$ ~~~~&$\cdots$~~~~&$\cdots$~~~~&$\cdots$~~~~&$\cdots$~~~~&$\cdots$~~~~     &$\cdots$~~~~\\
&$B_{c}(3F_3)$     ~~~~&$3^{+}$~~~~&7768        ~~~~~&7800    ~~~~&$\cdots$~~~~ &$\cdots$ ~~~~&$\cdots$~~~~&$\cdots$~~~~&$\cdots$~~~~&$\cdots$~~~~&$\cdots$~~~~     &$\cdots$~~~~\\
&$B_{c}(3 ^3F_{2})$~~~~&$2^{+}$~~~~&7730        ~~~~~&7800    ~~~~&7946    ~~~~ &$\cdots$ ~~~~&$\cdots$~~~~&$\cdots$~~~~&$\cdots$~~~~&$\cdots$~~~~&$\cdots$~~~~     &$\cdots$~~~~\\
\midrule[1.0pt]\midrule[1.0pt]
\end{tabular}
\end{table*}

\section{mass spectrum}\label{spectrum}


To describe a bottom-charmed meson system, we adopt
a nonrelativistic linear potential model. In this model, the effective quark-antiquark potential
is written as the sum of the spin-independent term $H_0(r)$ and spin-dependent term $H_{sd}(r)$, i.e.,
\begin{eqnarray}\label{H1}
V(r)=H_0(r)+H_{sd}(r),
\end{eqnarray}
where
\begin{eqnarray}\label{H0}
H_0(r)=-\frac{4}{3}\frac{\alpha_s}{r}+br
\end{eqnarray}
includes the standard color Coulomb interaction and the linear confinement.
The spin-dependent part $H_{sd}(r)$ can be expressed as~\cite{Eichten:1980mw,Kiselev:1994rc,Godfrey:2004ya}
\begin{eqnarray}\label{H0}
H_{sd}(r)=H_{SS}+H_{T}+H_{LS},
\end{eqnarray}
where
\begin{eqnarray}\label{ss}
H_{SS}= \frac{32\pi\alpha_s}{9m_qm_{\bar{q}}}\tilde{\delta}_\sigma(r)\mathbf{S}_{q}\cdot \mathbf{S}_{\bar{q}}
\end{eqnarray}
is the spin-spin contact hyperfine potential. Here, we take $\tilde{\delta}_\sigma(r)=(\sigma/\sqrt{\pi})^3
e^{-\sigma^2r^2}$ as suggested in Ref.~\cite{Barnes:2005pb}. The tensor potential $H_T$ is adopted as
\begin{eqnarray}\label{t}
H_{T}= \frac{4}{3}\frac{\alpha_s}{m_qm_{\bar{q}}}\frac{1}{r^3}\left(\frac{3\mathbf{S}_{q}\cdot \mathbf{r}\mathbf{S}_{\bar{q}}\cdot \mathbf{r}}{r^2}-\mathbf{S}_{q}\cdot\mathbf{S}_{\bar{q}}\right).
\end{eqnarray}
For convenience in the calculations, the potential of the spin-orbit interaction $H_{LS}$ is decomposed
into symmetric part $H_{sym}$ and antisymmetric part
$H_{anti}$,
\begin{eqnarray}\label{vs}
H_{LS}=H_{sym}+H_{anti},
\end{eqnarray}
with
\begin{eqnarray}\label{vs}
H_{sym}=\frac{\mathbf{S_{+}\cdot L}}{2}\left[\left(\frac{1}{2m_{\bar{q}}^{2}}+\frac{1}{2m_{q}^{2}}\right)\left(\frac{4}{3}\frac{\alpha_{s}}{r^{3}}-\frac{b}{r}\right)
+\frac{8\alpha_{s}}{3m_{q}m_{\bar{q}}r^{3}}\right],\\
H_{anti}=\frac{\mathbf{S_{-}\cdot L}}{2}\left(\frac{1}{2m_{q}^{2}}-\frac{1}{2m_{\bar{q}}^{2}}\right)\left(\frac{4}{3}\frac{\alpha_{s}}{r^{3}}-\frac{b}{r}\right).\ \ \ \ \ \ \ \ \ \ \ \ \ \ \ \ \ \ \ \ \ \ \
\end{eqnarray}
In these equations, $\mathbf{L}$ is the relative orbital angular momentum of the $q\bar{q}$
system; $\mathbf{S}_q$ and $\mathbf{S}_{\bar{q}}$ are the spins of the quark $q$ and antiquark $\bar{q}$, respectively, and $\mathbf{S}_{\pm}\equiv\mathbf{S}_q\pm \mathbf{S}_{\bar{q}}$; $m_q$ and $m_{\bar{q}}$ are the
masses of quark $q$ and antiquark $\bar{q}$, respectively; $\alpha_s$ is
the running coupling constant of QCD; and $r$ is the distance between the quark $q$ and antiquark $\bar{q}$.
The five parameters in the above equations ($\alpha_s$, $b$, $\sigma$, $m_b$, $m_c$) are determined by fitting the spectrum.

We can get the masses and wave functions by solving the radial Schr\"{o}dinger equation
\begin{eqnarray}\label{sdg}
\frac{d^2u(r)}{dr^2}+2\mu_R \left[E-V_{q\bar{q}}(r)-\frac{L(L+1)}{2\mu_R r^2}\right]u(r)=0,
\end{eqnarray}
with
\begin{eqnarray}\label{mtt}
V_{q\bar{q}}(r)=V(r)+H_{SS}+H_{SL}+H_{T},
\end{eqnarray}
where $\mu_R=m_q m_{\bar{q}}/(m_q+m_{\bar{q}})$ is the reduced mass
of the system, and $E$ is the binding energy of the system.
Then, the mass of a bottom-charmed state is obtained by
\begin{eqnarray}\label{mtt}
M_{q\bar{q}}=m_{q}+m_{\bar{q}}+E.
\end{eqnarray}

In this work, to reasonably include the corrections
from these spin-dependent potentials to both the mass and wave
function of a meson state, we deal with the spin-dependent
interactions nonperturbatively. We solve the radial Schr\"{o}dinger equation by using
the three-point difference central method~\cite{Haicai} from central ($r=0$)
towards outside ($r\to \infty$) point by point. This method was successfully to
deal with the spectroscopies of $c\bar{c}$ and $b\bar{b}$~\cite{Deng:2016ktl,Deng:2016stx}.  To overcome the singular
behavior of $1/r^3$ in the spin-dependent potentials, following
the method of our previous works~\cite{Deng:2016ktl,Deng:2016stx}, we introduce a cutoff distance
$r_c$ in the calculation. Within a small range $r\in (0,r_c)$, we let
$1/r^3=1/r_c^3$.

Finally, it should be mentioned that the antisymmetric part of the spin-orbit potential, $H_{anti}$,
can let the states with different total spins but with the same total angular momentum,
such as $B_{c}(n^{3}L_{J})$ and $B_{c}(n^{1}L_{J})$, mix with each other.
Thus, as mixing states between $B_{c}(n^{3}L_{J})$ and $B_{c}(n^{1}L_{J})$, the physical $B_c$ states
$B_{c}(nL)$ and $B_{c}(nL')$ are expressed as
\begin{equation}
\left(
  \begin{array}{c}
   B_{c}(nL'_J)\\
   B_{c}(nL_J)\\
  \end{array}\right)=
  \left(
  \begin{array}{cc}
   \cos\theta_{nL} &\sin\theta_{nL}\\
  -\sin\theta_{nL} &\cos\theta_{nL}\\
  \end{array}
\right)
\left(
  \begin{array}{c}
  B_{c}(n^{1}L_{J})\\
  B_{c}(n^{3}L_{J})\\
  \end{array}\right),
\end{equation}
where $J=L=1,2,3\cdots$, and the $\theta_{nL}$ is the mixing angle.
In this work $B_{c}(nL')$ corresponds to the higher mass mixed state
as often adopted in the literature.

\begin{figure*}[htbp]
\begin{center}
\centering  \epsfxsize=16.8cm \epsfbox{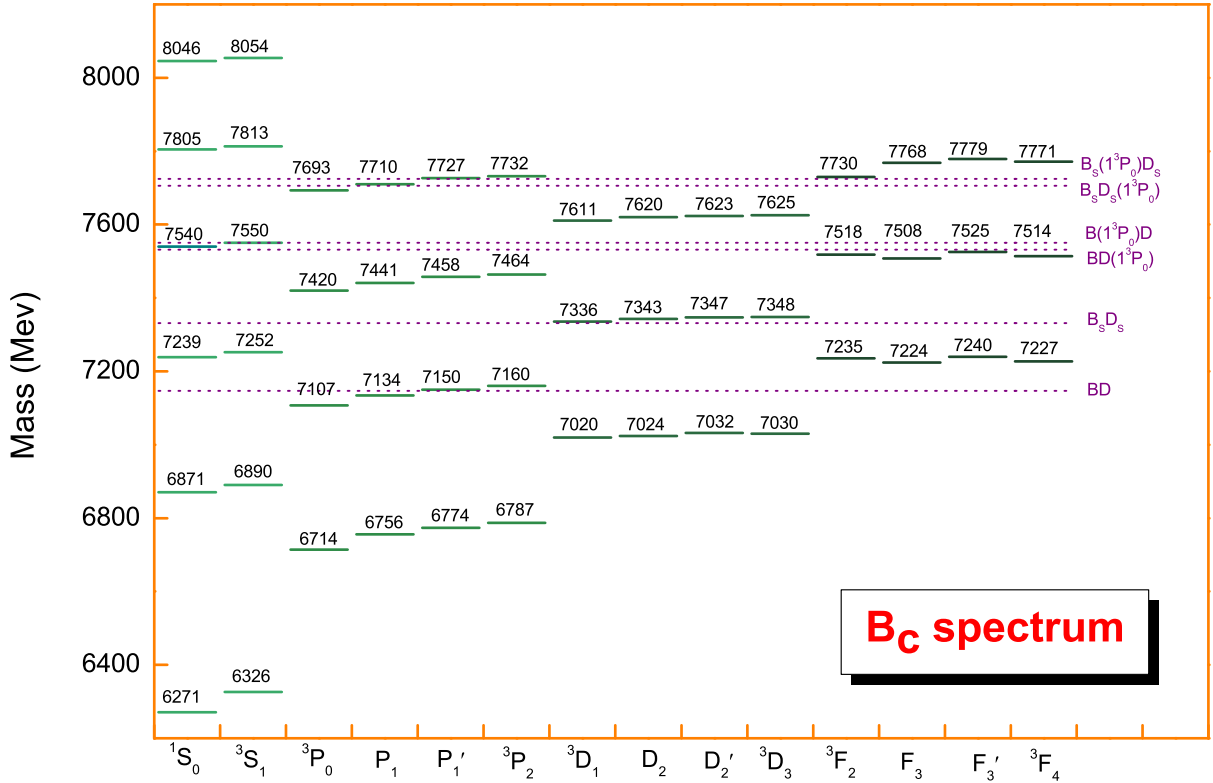}
\vspace{-1.0cm}\caption{The spectrum of $B_c$ mesons} \label{mass}
\end{center}
\end{figure*}


\begin{table}[htp]
\begin{center}
\caption{\label{mixangle} Mixing angles.
}{\begin{tabular}{ccccccccccccccccccccccccccccc}\hline\hline
~~~~~~~& Mixing angle  ~~~~~~~& ours    ~~~~~~~&\cite{Godfrey:2004ya}   ~~~~~~~&\cite{Devlani:2014nda}    ~~~~~~~&\cite{Ebert:2002pp}          \\
\hline
~~~~~~~& $\theta_{1P}$          ~~~~~~~& $35.5^{\circ}$ ~~~~~~~& $22.4^{\circ}$~~~~~~~& $20.57^{\circ}$~~~~~~~& $20.4^{\circ}$\\
~~~~~~~& $\theta_{2P}$          ~~~~~~~& $38.0^{\circ}$ ~~~~~~~& $18.9^{\circ}$~~~~~~~& $19.94^{\circ}$~~~~~~~& $23.2^{\circ}$\\
~~~~~~~& $\theta_{3P}$          ~~~~~~~& $39.7^{\circ}$ ~~~~~~~& $\cdots$      ~~~~~~~& $17.68^{\circ}$~~~~~~~& $\cdots$      \\
~~~~~~~& $\theta_{4P}$          ~~~~~~~& $39.7^{\circ}$  ~~~~~~~& $\cdots$      ~~~~~~~& $\cdots$       ~~~~~~~& $\cdots$      \\
~~~~~~~& $\theta_{1D}$          ~~~~~~~& $45.0^{\circ}$  ~~~~~~~& $44.5^{\circ}$~~~~~~~& $-2.49^{\circ}$~~~~~~~& $-35.9^{\circ}$\\
~~~~~~~& $\theta_{2D}$          ~~~~~~~& $45.0^{\circ}$  ~~~~~~~& $\cdots$      ~~~~~~~& $-2.8^{\circ}$ ~~~~~~~& $\cdots$      \\
~~~~~~~& $\theta_{3D}$          ~~~~~~~& $45.0^{\circ}$  ~~~~~~~& $\cdots$      ~~~~~~~& $\cdots$       ~~~~~~~& $\cdots$      \\
~~~~~~~& $\theta_{1F}$          ~~~~~~~& $41.4^{\circ}$  ~~~~~~~& $41.4^{\circ}$~~~~~~~& $\cdots$       ~~~~~~~& $\cdots$      \\
~~~~~~~& $\theta_{2F}$          ~~~~~~~& $43.4^{\circ}$ ~~~~~~~& $\cdots$      ~~~~~~~& $\cdots$       ~~~~~~~& $\cdots$      \\
~~~~~~~& $\theta_{3F}$          ~~~~~~~& $42.4^{\circ}$  ~~~~~~~& $\cdots$      ~~~~~~~& $\cdots$       ~~~~~~~& $\cdots$      \\
\hline\hline
\end{tabular}}
\end{center}
\end{table}

In this work the parameter set is taken as $\alpha_s=0.5021$, $b=0.1425$GeV$^2$, $m_b=4.852$ GeV,  $m_c=1.483$ GeV, $\sigma=1.3$ GeV and $r_c=0.16$ fm. To consistent with our previous study~\cite{Deng:2016stx}, the charmed quark mass $m_c$
and the slope for the linear confining potential are taken
the determinations, i.e., $m_c=1.483$ GeV and $b=0.1425$ GeV$^2$. The other three parameters
($m_b$, $\alpha_s$, $\sigma$) are determined by fitting the masses of the $B_{c}$, $B_{c}^{*}$ and $B_{c}(2S)$ mesons.
The masses of $B_{c}$ and $B_{c}(2S)$ are taken from the recent measurements of the CMS Collaboration~\cite{Sirunyan:2019osb}.
Although the $B_{c}^{*}$ meson is still not measured in experiments, the mass difference between the $B_c^*$ and $B_c$ is predicted to be around 55 MeV from lattice QCD~\cite{Gregory:2009hq,Dowdall:2012ab,Mathur:2018epb}. Thus, combining it with the measured
mass 6271 MeV for $B_{c}$, in present work we estimate the mass of $B_c^*$ as $\sim 6326$ MeV. The cutoff distance
$r_c$ is determined by the mass of $B_{c}(1 ^3P_{0})$. To determined the mass of $B_{c}(1 ^3P_{0})$, we adopt a method of perturbation, i.e., we let $H=H_{0}+H'$, where $H'$ is a part which contained the term of $1/r^{3}$. By solving the equation of $H_{0}|\psi_{n}^{(0)}\rangle=E_{0}|\psi_{n}^{(0)}\rangle$, we can get the energy $E_{0}$ and wave function $|\psi_{n}^{(0)}\rangle$, then, we obtain
the mass of $B_{c}(1 ^3P_{0})$, $M=m_b+m_c+E_{0}+\langle\psi_{n}^{(0)}|H'|\psi_{n}^{(0)}\rangle$.

By solving the radial Schr\"{o}dinger equation and with the determined parameter set,
we obtain the masses of the bottom-charmed states, which have been listed in Tab.~\ref{mass}
and shown in Fig.~\ref{mass}. For comparison, the other model predictions in Refs.~\cite{Zeng:1994vj,Eichten:1994gt,Ebert:2002pp,
Godfrey:2004ya,Kiselev:1994rc,Monteiro:2016ijw,Soni:2017wvy} are listed in the same table as well.

It is found that the masses of the low-lying $1S$-, $2S$-, $3S$-, $1P$-, $2P$-, $1D$-wave $B_c$ states predicted in this
work are compatible with the other potential model predictions.
For the higher mass states, such as $4S$-, $5S$-, $6S$-, $3P$-, $4P$-, $2D$-, $2F$-, $3F$-wave states,
the masses predicted by us are very close to those predicted with a relativistic model in Ref.~\cite{Zeng:1994vj},
while are about $100-200$ MeV smaller than those predicted in Refs.~\cite{Monteiro:2016ijw,Soni:2017wvy}.
Furthermore, the hyperfine splitting between $B_c^*(2S)$ and $B_c(2S)$ is predicted to be $19$ MeV,
which is slightly smaller than $30-45$ MeV predicted in previous works~\cite{Gregory:2009hq,Dowdall:2012ab,Mathur:2018epb,Zeng:1994vj,Soni:2017wvy,Monteiro:2016ijw,Eichten:1994gt,Ebert:2002pp,
Godfrey:2004ya,Kiselev:1994rc}. Finally, it should be pointed out the mixing angles for $^3P_1-^1P_1$, $^3D_2-^1D_2$, and $^3F_3-^1F_3$ have obvious model dependencies (see Tab.~\ref{mixangle}).

\begin{table*}[htb]
\caption{ Partial widths of the $M1$ transitions for the low-lying $1S$-, $2S$-, and $3S$-wave $B_c$ states compared with the other model predictions. }\label{M11}
\begin{tabular}{ccccccccccccc}  \midrule[1.0pt]\midrule[1.0pt]
 ~~~~~Initial~~~~~  & ~~~~~Final~~~~~& \multicolumn{5}{c} {\underline{~~~~~~~~~~~~~~~~~~~~~~~~~~~~$E_{\gamma}$  (MeV)~~~~~~~~~~~~~~~~~~~~~~~~~~~~}} & \multicolumn{4}{c} {\underline{~~~~~~~~~~~~~~~~~~~~~$\Gamma_{\mathrm{M1}}$  (eV)~~~~~~~~~~~~~~~~~~~~~}} & \multicolumn{2}{c} {\underline{$\Gamma_{\mathrm{M1}}$  (eV)}}  \\
   ~~~~~state~~~~~     & ~~~~~state~~~~~  &~~~~~\cite{Eichten:1994gt}~~~~~&~~~~~\cite{Ebert:2002pp}~~~~~&~~~~~\cite{Godfrey:2004ya}~~~~~& ~~~~~\cite{Kiselev:1994rc}~~~~~&~~~~~ours~~~~~
   &~~~~~\cite{Eichten:1994gt}~~~~~& ~~~~~\cite{Ebert:2002pp}~~~~~&~~~~~ \cite{Godfrey:2004ya}~~~~~
   & ~~~~~\cite{Kiselev:1994rc}~~~~~ &~~~~~ Ours~~~~~  										\\													
     \midrule[1.0pt]	          
$1 ^3S_{1}$	       &	$1 ^1S_{0}$   	&	72 	&	62      &	67	&	64	&	55	&	134.5	&	73  	&	80  	&	60  	&	57	    \\

$2 ^3S_{1}$	       &	$2 ^1S_{0}$   	&	43	&	46	    &	32	&	35	&	19	&	28.9	&	30  	&	10  	&	10  	&	2.4	    \\
        	       &	$1 ^1S_{0}$   	&	606	&	584	    &   588	&	649	&	591	&	123.4	&	141 	&	600 	&	98  	&	1205    \\
$2 ^1S_{0}$ 	   &	$1 ^3S_{1}$   	&	499	&	484	    &	498	&	550	&	523	&	93.3	&	160 	&	300 	&	96  	&	99      \\
$3 ^3S_{1}$	       &	$3 ^1S_{0}$   	&		&	   	    &	22	&	 	&	13	&	     	&	    	&	3   	&	    	&	0.8	    \\
        	       &	$2 ^1S_{0}$   	&		&	   	    &	405	&	 	&	371	&	     	&	    	&	200 	&	    	&	356	    \\
        	       &	$1 ^1S_{0}$   	&		&	   	    &	932	&	 	&	915	&	     	&	    	&	600 	&	    	&	1885    \\
$3 ^1S_{0}$	       &	$2 ^3S_{1}$   	&		&	   	    &	354	&	 	&	341	&	     	&	    	&	60  	&	    	&	152	    \\
        	       &	$1 ^3S_{1}$   	&		&	   	    &	855	&	 	&	855	&	     	&	    	&	4200	&	    	&	510     \\
\midrule[1.0pt]\midrule[1.0pt]
\end{tabular}
\end{table*}

\begin{table*}[htp]
\begin{center}
\caption{\label{radiative decay4} Partial widths of the $M1$ transitions for the higher $nS$-wave ($n=4,5,6$) $B_c$ states.}
\scalebox{1.0}{
\begin{tabular}{cccccccccccccccccccccccccccccccc}
\midrule[1.0pt]\midrule[1.0pt]
~~~& Initial state~~~~	&	Final state~~~~	    &	$E_{\gamma}$(MeV)	~~~~~~~~~~~&	$\Gamma_{\mathrm{EM}}$(eV)
~~~~~&	 Initial state~~~~	&	Final state~~~~	    &	$E_{\gamma}$(MeV)	~~~~~~~~~~~&	$\Gamma_{\mathrm{EM}}$(eV)	\\
\midrule[1.0pt]
~~~&  $4 ^1S_{0}$	~~~~~~~~& $3 ^3S_{1}$   ~~~~~~~~&	    283 	  ~~~~~~~~& 186    ~~~~~~~~&  $4 ^3S_{1}$ ~~~~~~~~&$4 ^1S_{0}$  ~~~~~~~~&	10             ~~&	0.35 \\
~~~&               ~~~~~~~~& $2 ^3S_{1}$   ~~~~~~~~&  	622 	    ~~~~~~~~& 579	   ~~~~~~~~&              ~~~~~~~~&$3 ^1S_{0}$  ~~~~~~~~&	 305	         ~~&   252 \\
~~~&        	      ~~~~~~~~& $1 ^3S_{1}$   ~~~~~~~~&  	1116	    ~~~~~~~~& 1122   ~~~~~~~~&          	  ~~~~~~~~&$2 ^1S_{0}$  ~~~~~~~~&	 648     ~~~~~~~~&   806 \\
~~~&               ~~~~~~~~&               ~~~~~~~~&             ~~~~~~~~&        ~~~~~~~~&              ~~~~~~~~&$1 ^1S_{0}$  ~~~~~~~~&	 1171    ~~~~~~~~&   2501\\
~~~&	$5 ^1S_{0}$	  ~~~~~~~~&	$4 ^3S_{1}$   ~~~~~~~~&		251       ~~~~~~~~&	209    ~~~~~~~~&	$5 ^3S_{1}$	~~~~~~~~&$5 ^1S_{0}$  ~~~~~~~~&	8 	     ~~~~~~~~&	0.18 \\
~~~&	              ~~~~~~~~&	$3 ^3S_{1}$   ~~~~~~~~&		533       ~~~~~~~~&	720    ~~~~~~~~&	        	  ~~~~~~~~&$4 ^1S_{0}$  ~~~~~~~~&	 268     ~~~~~~~~&	210  \\
~~~&	              ~~~~~~~~&	$2 ^3S_{1}$   ~~~~~~~~&		861       ~~~~~~~~&	1260   ~~~~~~~~&	            ~~~~~~~~&$3 ^1S_{0}$  ~~~~~~~~&	 553     ~~~~~~~~&	675  \\
~~~&	        	    ~~~~~~~~&	$1 ^3S_{1}$   ~~~~~~~~&		1339      ~~~~~~~~&	1893   ~~~~~~~~&	        	  ~~~~~~~~&$2 ^1S_{0}$  ~~~~~~~~&	 885     ~~~~~~~~&	1316 \\
~~~&	       	      ~~~~~~~~&	              ~~~~~~~~&		          ~~~~~~~~&	       ~~~~~~~~&	        	  ~~~~~~~~&$1 ^1S_{0}$  ~~~~~~~~&	 1390    ~~~~~~~~&	3107 \\
~~~&	$6 ^1S_{0}$	  ~~~~~~~~&	$5 ^3S_{1}$   ~~~~~~~~&		230       ~~~~~~~~&	225    ~~~~~~~~&	$6 ^3S_{1}$	~~~~~~~~&$6 ^1S_{0}$  ~~~~~~~~&	8        ~~~~~~~~&	0.18 \\
~~~&	              ~~~~~~~~&	$4 ^3S_{1}$   ~~~~~~~~&		481       ~~~~~~~~&	849    ~~~~~~~~&	            ~~~~~~~~&$5 ^1S_{0}$  ~~~~~~~~&	 245     ~~~~~~~~&	191  \\
~~~&	              ~~~~~~~~&	$3 ^3S_{1}$   ~~~~~~~~&		755       ~~~~~~~~&	1613   ~~~~~~~~&	         	  ~~~~~~~~&$4 ^1S_{0}$  ~~~~~~~~&	 498     ~~~~~~~~&	643  \\
~~~&	              ~~~~~~~~&	$2 ^3S_{1}$   ~~~~~~~~&		1073      ~~~~~~~~&	2203   ~~~~~~~~&	            ~~~~~~~~&$3 ^1S_{0}$  ~~~~~~~~&	 774     ~~~~~~~~&	1239 \\
~~~&	        	    ~~~~~~~~&	$1 ^3S_{1}$   ~~~~~~~~&		1536      ~~~~~~~~&	2822   ~~~~~~~~&	         	  ~~~~~~~~&$2 ^1S_{0}$  ~~~~~~~~&	 1096    ~~~~~~~~&	1917 \\
~~~&	              ~~~~~~~~&	              ~~~~~~~~&		          ~~~~~~~~&	       ~~~~~~~~&	            ~~~~~~~~&$1 ^1S_{0}$  ~~~~~~~~&	 1586    ~~~~~~~~&	3772 \\
\midrule[1.0pt]\midrule[1.0pt]
\end{tabular}}
\end{center}
\end{table*}

\section{RADIATIVE TRANSITIONS}\label{EMT}

We use the nonrelativistic constituent quark model as adopted in Refs.~\cite{Deng:2016ktl,Deng:2016stx,Wang:2017hej,Xiao:2017udy,Lu:2017meb,Wang:2017kfr,Yao:2018jmc}
to calculate the radiative transitions between the $B_c$ states.
In this model, the quark-photon EM coupling at the tree level
is taken as
\begin{eqnarray}\label{he}
H_e=-\sum_j
e_{j}\bar{\psi}_j\gamma^{j}_{\mu}A^{\mu}(\mathbf{k},\mathbf{r})\psi_j,
\end{eqnarray}
where $A^{\mu}$ represents the photon field with three momenta $\mathbf{k}$; while $e_j$ and $\mathbf{r}_j$
stand for the charge and coordinate of the constituent quark $\psi_j$, respectively.
In order to match the nonrelativistic wave functions of the $B_c$ states,
we adopt the nonrelativistic form of Eq.~(\ref{he}), which is given by ~\cite{Brodsky:1968ea,Li:1997gd,Zhao:2002id,Xiao:2015gra,Zhong:2011ti,Zhong:2011ht},
\begin{equation}\label{he2}
H_{e}^{nr}=\sum_{j}\left[e_{j}\mathbf{r}_{j}\cdot\veps-\frac{e_{j}}{2m_{j}
}\vsig_{j}\cdot(\veps\times\hat{\mathbf{k}})\right]e^{-i\textbf{k}\cdot
\textbf{r}_j},
\end{equation}
where $\boldsymbol \epsilon$ is the polarization vector of the final photon,
$m_j$ and $\vsig_j$ stand for the constituent mass and
Pauli spin vector for the $j$th quark. The helicity amplitude $\mathcal{A}$ can be expressed as
\begin{eqnarray}\label{amp3}
\mathcal{A}&=&-i\sqrt{\frac{\omega_\gamma}{2}}\langle f | H_{e}^{nr}| i
\rangle.
\end{eqnarray}
Finally, we obtain the partial decay width of a radiative transition by
\begin{equation}\label{dww}
\Gamma=\frac{|\mathbf{k}|^2}{\pi}\frac{2}{2J_i+1}\frac{M_{f}}{M_{i}}\sum_{J_{fz},J_{iz}}|\mathcal{A}_{J_{fz},J_{iz}}|^2,
\end{equation}
where $J_i$ is the total angular momentum of an initial meson, and
$J_{fz}$ and $J_{iz}$ are the components of the total angular
momenta along the $z$ axis of initial and final mesons, respectively.
$M_i$ and $M_f$ correspond to the  masses of the initial and final $B_c$ states, respectively.

The radiative decays properties for the $B_c$ states have been listed in Tables ~\ref{M11}-~\ref{radiative decay5}.
For a comparison, some other predictions of the low-lying $B_c$ states from Refs.~\cite{Eichten:1994gt,Ebert:2002pp,Godfrey:2004ya,Kiselev:1994rc} are also given in the tables.

\section{strong decays}\label{Strongdecay}

In this work, we use the $^3P_0$ model~\cite{Micu:1968mk,LeYaouanc:1972vsx,LeYaouanc:1973ldf} to calculate the OZI-allowed strong decays of the bottom-charmed mesons. In this model, it assumes that the vacuum produces a quark-antiquark pair with the quantum
number $0^{++}$ and the heavy meson decay takes place via the rearrangement of the four quarks.
The transition operator $\hat{T}$ in this model can be written as
\begin{eqnarray}
    \hat{T} & = & -3 \gamma \sqrt{96 \pi} \sum_{m}^{} \langle 1 m 1 -m| 0 0 \rangle \int_{}^{} d\mathbf{p}_3 d\mathbf{p}_4 \delta^3 (\mathbf{p}_3 + \mathbf{p}_4) \nonumber\\
      & \times &  \mathcal{Y}_1^m \left(\frac{\mathbf{p}_3 - \mathbf{p}_4}{2}\right)  \chi_{1-m}^{34}  \phi_0^{34} \omega_0^{34} b_{3i}^\dagger (\mathbf{p}_3) d_{4j}^\dagger (\mathbf{p}_4) \ ,
\end{eqnarray}
where $\gamma$ is a dimensionless constant that denotes the strength of the quark-antiquark pair creation with
momentum $\mathbf{p}_3$ and $\mathbf{p}_4$ from vacuum; $b_{3i}^\dagger (\mathbf{p}_3)$ and $d_{4j}^\dagger(\mathbf{p}_4)$ are the creation operators for the quark and antiquark, respectively; the subscriptions, $i$ and $j$, are the SU(3)-color indices of the created quark and anti-quark;
$\phi_0^{34}=(u\bar u +d\bar d +s \bar s)/\sqrt 3$ and $\omega_{0}^{34}=\frac{1}{\sqrt{3}} \delta_{ij}$ correspond to flavor and
color singlets, respectively; $\chi_{{1,-m}}^{34}$ is a spin triplet
state; and $\mathcal{Y}_{\ell m}(\mathbf{k})\equiv
|\mathbf{k}|^{\ell}Y_{\ell m}(\theta_{\mathbf{k}},\phi_{\mathbf{k}})$ is the
$\ell$-th solid harmonic polynomial. The factor $(-3)$ is introduced for
convenience, which will cancel the color factor.

For an OZI allowed two-body strong decay process $A\to B+C$, the helicity amplitude
$\mathcal{M}^{M_{J_A}M_{J_B} M_{J_C}}(\mathbf{P})$ can be derived as follow
\begin{eqnarray}
\langle BC|T| A\rangle=\delta(\mathbf{P}_A-\mathbf{P}_B-\mathbf{P}_C)\mathcal{M}^{M_{J_A}M_{J_B} M_{J_C}}(\mathbf{P}).
\end{eqnarray}
Using the Jacob-Wick formula~\cite{Jacob:1959at},
one can convert the helicity amplitudes
$\mathcal{M}^{M_{J_A}M_{J_B} M_{J_C}}(\mathbf{P})$ to the partial wave amplitudes $\mathcal{M}^{JL}$ via
\begin{equation}\label{eq4}
\begin{aligned}
&
{\mathcal{M}}^{J L}(A\rightarrow BC) = \frac{\sqrt{4\pi (2 L+1)}}{2 J_A+1} \!\! \sum_{M_{J_B},M_{J_C}} \langle L 0 J M_{J_A}|J_A M_{J_A}\rangle \\
& \hspace{1cm}
\times  \langle J_B M_{J_B} J_C M_{J_C} | J M_{J_A} \rangle \mathcal{M}^{M_{J_A} M_{J_B} M_{J_C}}({\textbf{P}}).
\end{aligned}
\end{equation}
In the above equations, ($J_{A}$, $J_{B}$ and $J_{C}$), ($L_A$, $L_B$ and $L_C$) and ($S_A$, $S_B$ and $S_C$) are the quantum numbers of the total angular momenta, orbital angular momenta and total spin for hadrons $A,B,C$, respectively; $M_{J_A}=M_{J_B}+M_{J_C}$ ,\;$\mathbf{J}\equiv \mathbf{J}_B+\mathbf{J}_C$ and $\mathbf{J}_{A} \equiv \mathbf{J}_{B}+\mathbf{J}_C+\mathbf{L}$.
In the c.m. frame of hadron $A$, the momenta $\mathbf{P}_B$ and $\mathbf{P}_C$ of mesons $B$ and $C$ satisfy
$\mathbf{P}_B=-\mathbf{P}_C\equiv \mathbf{P}$.

Then the strong decay partial width for a given decay mode of
$A\to B+C$ is given by
\begin{eqnarray}
\Gamma = 2\pi |\textbf{P}| \frac{E_BE_C}{M_A}\sum_{JL}\Big
|\mathcal{M}^{J L}\Big|^2,\label{de}
\end{eqnarray}
where $M_A$ is the mass of the initial hadron $A$, while $E_B$ and $E_C$ stand for the energies of
final hadrons $B$ and $C$, respectively. The details of the $ ^{3}P_{0} $ model can be found in our recent paper~\cite{Gui:2018rvv}.

In the calculations, the wavefunctions of the initial $B_c$ states are adopted our quark model predictions.
Furthermore, we need the wavefunctions of the final hadrons, i.e.,
the $B^{(*)}$, $B_s^{(*)}$, $D^{(*)}$, $D_s^{(*)}$ mesons and some of their excitations,
which are adopted from the quark model predictions of Refs.~\cite{Lu:2016bbk,Li:2010vx}.

\begin{table}[htp]
\begin{center}
\caption{\label{meson mass} The masses (MeV) of the final hadrons appearing in the strong decay processes of the $B_c$ states.
The masses are taken from the Particle Data Group~\cite{Tanabashi:2018oca} if there are experimental data, otherwise we take the quark model predictions in Refs.~\cite{Lu:2016bbk,Li:2010vx}.
}{\begin{tabular}{ccccccccccccccccc}\hline\hline
~~~~& State     ~~~~& $1 ^1S_{0}$  ~~~~& $1 ^3S_{1}$  ~~~~& $1 ^3P_{0}$  ~~~~& $1P_1$          ~~~~& $1P'_1$       ~~~~& $1 ^3P_{2}$~~~~ \\\hline
~~~~& $B$       ~~~~& 5279         ~~~~& 5325         ~~~~& 5683         ~~~~& 5729          ~~~~& 5754       ~~~~& 5768       ~~~~ \\
~~~~& $B_{s}$   ~~~~& 5367         ~~~~& 5415         ~~~~& 5756         ~~~~& 5801          ~~~~& 5836       ~~~~& 5851       ~~~~ \\
~~~~& $D$       ~~~~& 1870         ~~~~& 2010         ~~~~& 2252         ~~~~& 2402          ~~~~& 2417       ~~~~& 2466       ~~~~ \\
~~~~& $D_{s}$   ~~~~& 1968         ~~~~& 2112         ~~~~& 2344         ~~~~& 2488          ~~~~& 2510       ~~~~& 2559       ~~~~ \\
\hline\hline
\end{tabular}}
\end{center}
\end{table}

In this work, for the masses of the light constituent $u$, $d$ and $s$ quarks, we set $m_u=m_d=0.33$ GeV, $m_s=0.45$GeV;
while for the heavy $b$ and $c$ quarks, their masses are taken to be $m_b=4.852$ GeV and $m_c=1.483$ GeV
as the determinations in the calculations of the $B_c$ mass spectrum. The masses of the final hadron states in
the decay processes are adopted from the Particle Data Group~\cite{Tanabashi:2018oca} if there are measured values, otherwise we take the quark model predictions of Refs.~\cite{Lu:2016bbk,Li:2010vx} (see Table~\ref{meson mass}). There is no experimental data which
can be used to determine the quark pair creation strength, thus, in this work we adopt a typical value $\gamma=0.4$ that gives a reasonably accurate description of the overall scale of decay widths of both light and heavy mesons~\cite{Ackleh:1996yt,Godfrey:2016nwn,Barnes:2002mu,Barnes:2005pb,Godfrey:2015dva,Close:2005se}.
The strong decays properties for the bottom-charmed states
are presented in Tab.~\ref{strong decay5} to ~\ref{strong decay10}.

\section{discussion}\label{DIS}

\subsubsection{$S$-wave states}

Recently, signals of two excited $\bar{b}c$ states $B_c(2S)$ and
$B^*_c(2S)$ were observed in the $B_c^+\pi^+\pi^-$ invariant
mass spectrum by the CMS Collaboration at LHC. These two states are well resolved from each other and are observed with a significance
exceeding five standard deviations. The mass of $B_c(2S)$ meson is measured to be $6871\pm 2.8$ MeV.
Furthermore, a more precise mass of $B_c(2S)$, $M(B_c^+)=6871.1\pm 0.5$ MeV, is measured by the CMS Collaboration as well.
Combining these newest measurements, we predict that the mass of $B_c(2S)$ might be $\sim 6890$ MeV,
and the mass hyperfine splitting between $B^*_c(2S)$ and $B_c(2S)$,
\begin{eqnarray}
\Delta m(2S)\simeq 20 \ \ \mathrm{MeV},
\end{eqnarray}
is slightly smaller than $30-45$ MeV predicted in previous works (see Table~\ref{mass}).
The predicted masses for the other higher $S$-wave states compared with other works are also
given in Table~\ref{mass}. Obvious differences can be found in various theoretical predictions.

The $M1$ transitions of the low-lying $S$-wave states $B_c^*(2S)$ and $B_c^{(*)}(1S)$ were often
discussed in the literature for these transitions which might be used to establish them in experiments. In this work we also calculate their $M1$ transitions. Our results compared with the some other predictions are listed Table~\ref{M11}.
Obvious model dependence can be seen in various calculations. Our predicted partial width
\begin{eqnarray}
\Gamma[B_c^*(2S)\to B_c \gamma]\simeq 1.2\ \ \mathrm{keV},
\end{eqnarray}
for the $M1$ transition $B_c^*(2S)\to B_c \gamma$ is about an order of magnitude
larger than that predicted in Refs.~\cite{Eichten:1994gt,Ebert:2002pp,Kiselev:1994rc}, and about a factor 2
larger than the value predicted within the GI model~\cite{Godfrey:2004ya}. Combining our calculations of the EM transitions $B_c^*(2S)\to 1 P \gamma$ and
the strong transitions $B_c^*(2S)\to B_c^* \pi\pi$ predicted in~\cite{Godfrey:2004ya}, the total decay width of $B_c^*(2S)$
meson is estimated to be $\Gamma_{total}\sim 75$ keV, then the branching fraction for
$M1$ transition $B_c^*(2S)\to B_c \gamma$ is predicted to be
\begin{eqnarray}
Br[B_c^*(2S)\to B_c \gamma]\sim 2\%.
\end{eqnarray}
The fairly large branching fraction may give a good opportunity for us
to observe the $B_c^*(2S)$ via the $M1$ transition $B_c^*(2S)\to B_c \gamma$. This process
may be used to determined the mass of $B_c^*(2S)$ in future experiments.

The masses of $3S$-wave states $B_c(3^1S_0)$ and $B_c(3^3S_1)$ are predicted to be $\sim7.24$ GeV
and $\sim7.25$ GeV, respectively, which are just above the $DB^*$ threshold.
Their radiative and strong decay properties are estimated in this work. The results for the $M1$ transitions, $E1$ dominant transitions and strong decays of the $3S$-wave states are given in Tables~\ref{M11},~\ref{radiative decay5} and~\ref{strong decay5}, respectively.
There are only a few works about the radiative and strong decay properties of the $3S$-wave states~\cite{Godfrey:2004ya,Kiselev:1996un,Ferretti:2015rsa,Eichten:2019gig}. The $M1$ transitions of the $3S$-wave states roughly agree with the predictions in Ref.~\cite{Godfrey:2004ya}, except that our predicted partial width
$\Gamma[3^3S_1\to 1^1S_0+\gamma]\simeq 510$ eV for the $M1$ transition $3^3S_1\to 1^1S_0+\gamma$ is about an order of magnitude smaller than that in Ref.~\cite{Godfrey:2004ya}. The strong decay widths of $B_c(3^1S_0)$ and $B_c(3^3S_1)$ predicted by us are comparable to those predicted
in recent works~\cite{Ferretti:2015rsa,Eichten:2019gig}.
Both $B_c(3^1S_0)$ and $B_c(3^3S_1)$  might be broad states with a width of $\sim 100$ MeV.
The $B_c(3^1S_0)$ dominantly decay into $DB^*$ channel, while $B_c(3^3S_1)$ dominantly decay into
both $DB$ and $DB^*$ channels. The production rates of the $3S$-wave $B_c$ states in $pp$
collisions at the LHC may be comparable with those of the $2S$-wave $B_c$ states~\cite{Eichten:2019gig},
thus, the $3S$-wave $B_c$ states may have large potentials to be established in the $DB^*$ final states.

The higher $S$-wave states $B_c(n^1S_0)$ and $B_c(n^3S_1)$ ($n\geq 4$) are far from the $DB$ threshold, thus
many OZI-allowed two-body strong decay channels are open. There are few discussions of the decay properties of the higher
mass $S$-wave states in the literature. To know some decay properties of these higher $S$-wave states, in this work we give our predictions of the $M1$ transitions and strong decays of $B_c(nS)$ ($n=4,5,6$), which are listed in Table~\ref{radiative decay4} and ~\ref{strong decay5}, respectively. It is found these higher mass $S$-wave states are broad states with a width of $\sim 100-400$ MeV.
Combining $M1$ transitions of higher $S$-wave states with their strong decays, we found that the branching fractions of the
$M1$ transitions $B_c(nS)\to B_c(1S)+\gamma$ may reach up to a sizeable value $\mathcal{O}(10^{-5})$.

\subsubsection{$P$-wave states}

The masses of $1P$-wave states $B_c(1P)$ might lie in the range of $(6710,~6790)$ MeV,
which are consistent with the other predictions with potential models~\cite{Eichten:1994gt,Ebert:2002pp,
Zeng:1994vj,Godfrey:2004ya,Kiselev:1994rc}, and the recent lattice calculations~\cite{Mathur:2018epb}.
The $1P$-wave $B_c(1P)$ states mainly decays via the $E1$ dominate transitions $1P\to 1S$.
We have calculated the partial decay widths for the $EM$ transitions $1P\to 1S$, our results
compared with some other predictions are listed in Table~\ref{radiative decay3}.
Most of our results are compatible with the predictions in~\cite{Eichten:1994gt,Ebert:2002pp,
Godfrey:2004ya,Kiselev:1994rc}, except our predicted partial decay widths of $\Gamma[B_c(1P_1)\to B_c\gamma]\simeq 35$ keV
and $\Gamma[B_c(1P_1')\to B_c^*\gamma]\simeq 40$ keV are about a factor of $3-5$ larger than the predictions in Refs.~\cite{Ebert:2002pp,
Godfrey:2004ya,Kiselev:1994rc}. The $B_c(1P_1)$ and $B_c(1P_1')$ states might be first found in the $B_c\gamma$ final state via their radiative transitions. The branching fractions for $B_c(1P_1)$ and $B_c(1P_1')$  decay into $B_c\gamma$ are predicted to be
\begin{eqnarray}
Br[B_c(1P_1)\to B_c\gamma]\sim 33\%,\\
Br[B_c(1P_1')\to B_c\gamma]\sim 65\%.
\end{eqnarray}
While the $B_c(1^3P_0)$ and $B_c(1^3P_2)$ states dominantly decay into $B_c^*\gamma$ final state with a decay rate $\sim100\%$,
thus, they have good potentials to be found via the radiative decay chains $B_c(1^3P_0)\to B_c(1^3S_1)\gamma\to B_c(1^1S_0)\gamma\gamma$ and $B_c(1^3P_2)\to B_c(1^3S_1)\gamma\to B_c(1^1S_0)\gamma\gamma$, respectively.

For the $2P$-wave states $B_c(2P)$, their masses might lie in the range $(7100,~7160)$ MeV,
which are consistent with the other model predictions in the literature~\cite{Zeng:1994vj,Eichten:1994gt,Ebert:2002pp,
Godfrey:2004ya,Kiselev:1994rc,Soni:2017wvy,Monteiro:2016ijw}. The masses for $B_c(2^3P_0)$ and $B_c(2P_1)$
are slightly lower than the $DB$ mass threshold, while $B_c(2P'_1)$ and $B_c(2^3P_2)$ slightly lie above the
$DB$ mass threshold. The $B_c(2^3P_2)$ state mainly decay into the $DB$ channel, while its radiative decay rates into the $B_c(n^3S_1)\gamma$ ($n=1,2$) are also sizeable. Their partial widths are predicted to be
\begin{eqnarray}
&&\Gamma[B_c(2^3P_2)\to DB]\simeq 760 ~\mathrm{keV},\\
&&\Gamma[B_c(2^3P_2)\to B_c^*\gamma]\simeq 52 ~\mathrm{keV},\\
&&\Gamma[B_c(2^3P_2)\to B_c^*(2S)\gamma]\simeq 50 ~\mathrm{keV},
\end{eqnarray}
Thus, the total width of $B_c(2^3P_2)$ is $\Gamma_{\mathrm{total}}[B_c(2^3P_2)]\simeq 880$ keV.
The $B_c(2^3P_2)$ state may have potentials to be observed in the $DB$ and $B_c\gamma$ final states.
While for $B_c(2^3P_0)$, $B_c(2P_1)$ and $B_c(2P'_1)$ states, their decays are governed by the $EM$ transitions.
The radiative decay properties of these states have been given in Table~\ref{radiative decay2}.
With these predictions, the total widths for $B_c(2^3P_0)$, $B_c(2P_1)$ and $B_c(2P'_1)$ are estimated to
be $\Gamma_{\mathrm{total}}[B_c(2^3P_0)]\simeq 100$ keV, $\Gamma_{\mathrm{total}}[B_c(2P_1)]\simeq 120$ keV,
and  $\Gamma_{\mathrm{total}}[B_c(2P_1')]\simeq 133$ keV, respectively. The branching fractions for
$B_c(2P_1)\to B_c\gamma$, $B_c(2P'_1)\to B_c\gamma$ and $B_c(2^3P_0)\to B_c^*\gamma$ are predicted to be
\begin{eqnarray}
&&Br[B_c(2P_1)\to B_c\gamma]\simeq 20\%,\\
&&Br[B_c(2P_1')\to B_c\gamma]\simeq 33\%\\
&&Br[B_c(2^3P_0)\to B_c^*\gamma]\simeq 41\%.
\end{eqnarray}
The large branching fractions indicate that $B_c(2P_1)$ and $B_c(2P'_1)$
may be established in the $B_c\gamma$ channel, while $B_c(2^3P_0)$ may be observed via the radiative decay chain
$B_c(2^3P_0)\to B_c^*\gamma\to B_c\gamma\gamma$. It should be pointed out that the $B_c(2P_1)$, $B_c(2P'_1)$ and $B_c(2^3P_2)$
states may lie above the $B^*D$ threshold, so they may have fairly large strong decay widths $\mathcal{O}(10-100)$ MeV into $B^*D$ and/or $BD$ channels as predicted in Ref.~\cite{Monteiro:2016rzi}.

For the higher $P$-wave states $B_c(nP)$ ($n=3,4$), many OZI allowed strong decay
channels are open (see Table \ref{strong decay7}), thus, these states usually are broad
states with a width of $\mathcal{O}(100)$ MeV, except the $B_c(4^3P_0)$ state has a
relatively narrow width of $\mathcal{O}(10)$ MeV. The $B_c(4^3P_0)$ state may be first observed in
the $DB$ channel, the branching fraction for the process $B_c(4^3P_0)\to DB$ can reach up to $\sim20\%$.

\subsubsection{$D$-wave states}

The masses of the $1D$-wave states $B_c(1D)$ is predicted to be $\sim 7.02$ GeV in this work. The mass
splitting between the $1D$-wave states is no more than $15$ MeV. The masses predicted by us
are consistent with the results in Refs.~\cite{Zeng:1994vj,Eichten:1994gt,
Godfrey:2004ya}. The $1D$-wave states mainly decay via the $EM$ transitions,
which have been given in Table~\ref{radiative decay3}. It is seen that our main results are in reasonable agreement
with the other predictions. Our study indicates that the $B_c(1^3D_3)$ state may have a relatively large potential
to be observed via the radiative decay chain
$B_c(1^3D_3)\to B_c(1^3P_2)\gamma\to B_c(1^3S_1)\gamma\gamma\to B_c(1^1S_0)\gamma\gamma\gamma$,
and the branching fraction for this chain is estimated to be $\sim100\%$.
The optimal decay chain for the observations of $B_c(1^3D_1)$ is $B_c(1^3D_1)\to B_c(1^3P_0)\gamma\to B_c(1^3S_1)\gamma\gamma\to B_c(1^1S_0)\gamma\gamma\gamma$, and the branching fraction for this chain is estimated to be $\sim60\%$.
The optimal decay chains for the observations of $B_c(1D_2)$ are $B_c(1D_2)\to B_c(1P_1)\gamma\to B_c(1^3S_1)\gamma\gamma\to B_c(1^1S_0)\gamma\gamma\gamma$ and $B_c(1D_2)\to B_c(1P_1)\gamma\to B_c(1^1S_0)\gamma\gamma$,
and the branching fraction for these chains are estimated to be $\sim50\%$ and $\sim30\%$, respectively.
While for the observations of $B_c(1D_2')$, the optimal decay chains are $B_c(1D_2')\to B_c(1P_1')\gamma\to B_c(1^3S_1)\gamma\gamma\to B_c(1^1S_0)\gamma\gamma\gamma$ and $B_c(1D_2')\to B_c(1P_1')\gamma\to B_c(1^1S_0)\gamma\gamma$,
and the branching fraction for these chains are estimated to be $\sim35\%$ and $\sim47\%$, respectively.

The masses of the $2D$ states are predicted to be $\sim7.34$ GeV, which is very close to the $D_sB_s$ threshold.
Their decays are governed by the strong decay modes, such as $DB$, $DB^*$, $BD^*$ or $B^*D^*$.
Their strong decay properties predicted by us have been listed in Table~\ref{strong decay8}.
There are few discussions about the radiative decays of the
$2D$-wave $B_c$ states in the literature. In this work, we also calculate their radiative decay properties,
our results are given in Table~\ref{radiative decay5}.
It is found that the $B_c(2^3D_1)$ state has a relatively narrow width of $\Gamma\sim 58$ MeV. The decays of
$B_c(2^3D_1)$ are governed by the $BD^*$ mode with a branching fraction
\begin{eqnarray}
Br[B_c(2^3D_1)\to BD^*]\simeq 87\%.
\end{eqnarray}
The other three $2D$ states $B_c(2^3D_3)$, $B_c(2D_2)$ and
$B_c(2D_2')$ are broad states with a width of $\sim 100-200$ MeV. The $B_c(2^3D_3)$ state mainly decays into
$DB$, $DB^*$, and $B^*D^*$ channels. While the $B_c(2D_2)$ and $B_c(2D_2')$ states dominantly decay
into $DB^*$, $BD^*$ or $B^*D^*$ channels. Combing the strong and radiative decay properties with each other,
it is found that the branching fractions of the dominant $EM$ decay processes $B_c(2D)\to B_c(nP)$ ($n=1,2$)
are $\mathcal{O}(10^{-4})$. The observations of the $DB$, $DB^*$, $BD^*$ or $B^*D^*$
final states might be useful to search for these missing $2D$ states in future experiments.

The higher $3D$-wave states $B_c(3D)$ are also studied in present work. The masses predicted by us are
about $7.62$ GeV, which are comparable with those predicted in Ref.~\cite{Zeng:1994vj},
while about 150 MeV smaller than those predicted in Refs.~\cite{Soni:2017wvy,Monteiro:2016ijw}.
The strong decay properties are shown in Table~\ref{strong decay8}.
It is found that these higher $3D$-wave states have a width of $\sim 100$ MeV.
These higher states might be observed in their dominant strong decay channels.

\subsubsection{$F$-wave states}

The masses of the $1F$-wave states $B_c(1^3F_4)$, $B_c(1F_3)$, $B_c(1F_3')$
and $B_c(1^3F_2)$ are predicted to be $\sim 7.23$ GeV,
which are comparable to those predicted in Refs.~\cite{Zeng:1994vj,Soni:2017wvy,Godfrey:2004ya}.
These $1F$ wave states lie above the mass threshold of $DB$ and $B^*D$, while below the $D^*B$ threshold.
From our predictions of the strong decay properties for these $1F$ wave states (see Table~\ref{strong decay9}),
it is found that the $B_c(1^3F_4)$ state might be a very narrow state with a width of $\sim 1$ MeV,
its decays are governed by the $DB$ mode. Both $B_c(1F_3)$ and $B_c(1F_3')$ are narrow states with a
width of $\sim 10$ MeV, they dominantly decay into the $DB^*$ channel. The $B_c(1^3F_2)$ should be a relatively broad
state with a width of $\sim 70$ MeV, it mainly decays into the $DB$ channel with a branching fraction of
$Br[B_c(1^3F_2)\to DB]\simeq 85\%$. To look for the missing $1F$-wave $B_c$ states,
the $DB$ and $B^*D$ final states are worth observing.

The predicted masses for the $2F$- and $3F$-wave $B_c$ states are $\sim7.5$ GeV and $\sim7.8$ GeV, respectively,
which are comparable with the predictions in Refs.~\cite{Zeng:1994vj,Godfrey:2004ya}. There are many strong
decay channels for these higher mass $F$-wave states. Our predictions of their strong decay properties
have been listed in Tables~\ref{strong decay9} and ~\ref{strong decay10}. It is found that the higher mass $F$-wave states
might be broad states with a width of $\sim 100-300$ MeV.

\section{summary}\label{sum}

In this paper, we have calculated the $B_c$ meson spectrum up to the $6S$ states with a nonrelativistic linear potential model by further constraining the model parameters with the mass of $B_c(2S)$ newly measured by the CMS Collaboration. As important tasks of this work, the radiative transitions between the $B_c$ states and the OZI allowed two body strong decays for the higher mass excited $B_c$ states are evaluated with the wavefunctions obtained from the linear potential model. Our calculations may provide useful references to search for the excited $B_c$ states. The main results are emphasized as follows.

For the $S$-wave states, the $2S$ hyperfine splitting is predicted to be $m[B_c^*(2S)]-m[B_c(2S)]\simeq 19$ MeV. The mass of the newly observed $B_c^*(2S)$ state might be determined via the $M1$ transition $B_c^*(2S)\to B_c\gamma$ in future experiments. The $3S$-wave states $B_c(3^1S_0)$ and $B_c(3^3S_1)$ are about $50$ MeV above the $DB^*$ threshold, their widths are estimated to be $\sim 100$ MeV. Since production rates of the $3S$-wave $B_c$ states in $pp$
collisions at the LHC are comparable with those of the $2S$-wave $B_c$ states~\cite{Eichten:2019gig},
both $B_c(3^1S_0)$ and $B_c(3^3S_1)$ states may have large possibilities to be established in the $DB^*$ final state,
while $B_c(3^3S_1)$ might be observed in the $DB$ final state as well.

For the $P$-wave states, it is found that the decays of the $2P$-wave states, $B_c(2^3P_0)$, $B_c(2P_1)$ and $B_c(2P'_1)$ together with all of the $1P$-wave states are governed by the $E1$ transitions, their typical decay widths are $\sim 100$ keV. It should be possible to observe these $P$-wave states via their dominant radiative decay processes with the higher statistics of the LHC.
The $B_c(2^3P_2)$ state is just $\sim 20$ MeV above the $DB$ threshold. It mainly decays into the $DB$ channel
with a very narrow width of $\Gamma\sim 1$ MeV, so it has a large potential to be first observed in the $DB$ final state.
The predicted masses of $3P$-wave states are in the range of (7420,7470) MeV. They are broad states with widths of $\sim 200$ MeV, and strongly couple to the $B^*D^*$ final state. It is interesting found that the $4P$-wave states $B_c(4^3P_0)$, $B_c(4P_1)$ and $B_c(4P'_1)$ with a mass around $7.7$ GeV may have relatively narrow widths $\mathcal{O}(100)$ MeV, these higher $P$-wave states might be first observed in their dominant channel $DB$ or $DB^*$.

The $1D$-wave states mainly decay via the $EM$ transitions. Our study indicates that these $1D$-wave states may have a relatively large potential
to be observed via the radiative decay chains. For example, to look for the $B_c(1^3D_3)$ state, the $B_c(1^3D_3)\to B_c(1^3P_2)\gamma\to B_c(1^3S_1)\gamma\gamma\to B_c(1^1S_0)\gamma\gamma\gamma$ is worthy to be searched, for the branching fraction of this chain is estimated to be $\sim100\%$. The masses of the $2D$ and $3D$ states are predicted to be $\sim7.34$ and $7.62$ GeV, respectively. Their decays are governed by the strong decay modes, such as $DB$, $DB^*$, $BD^*$ or $B^*D^*$. These higher $D$-wave states usually have a width of $\mathcal{O}(100)$ MeV.
The observations of the $DB$, $DB^*$, $BD^*$ or $B^*D^*$ final states might be useful to search for these missing $2D$ and $3D$ states in future experiments.

For the $F$-wave states, one should pay more attention to $1F$-wave $B_c$ states in future observations.
They have a mass of $\sim 7.23$ GeV, lie between the $DB$ and $B^*D$ mass thresholds. They are narrow states
with a width of several MeV to several ten MeV, and dominantly decay into $DB$ or $B^*D$ channels. For example the
$B_c(1^3F_4)$ state might be a very narrow state with a width of $\sim 1$ MeV, its decays are governed by the $DB$ mode.
To look for the missing $1F$-wave $B_c$ states, the $DB$ and $B^*D$ final states are worth observing.

Finally, it should be pointed out the strong decay widths of the excited $B_c$ states predicted in this work
may have large uncertainties, for the parameter $\gamma$ cannot be directly determined by the strong decay processes of $B_c$ states.
Fortunately, the uncertainties of the total strong decay widths of the excited $B_c$ states do not affect the important information, such as the dominant decay modes and corresponding decay rates, for our searching for the excited $B_c$ states in future experiments.
Furthermore, the mixing angles for $^3P_1-^1P_1$, $^3D_2-^1D_2$, and $^3F_3-^1F_3$ have obvious model dependencies.
The uncertainties of the mixing angles also affect our predictions of the decay properties of the mixed states.

\begin{table*}[htb]
\caption{Partial widths of the $E1$ dominant radiative transitions for the $1P$-, $1D$-, and $1F$-wave $B_c$ states.
For comparison, the predictions from the relativistic quark model~\cite{Ebert:2002pp}, relativized quark model~\cite{Godfrey:2004ya}, nonrelativistic constituent quark models~\cite{Eichten:1994gt,Kiselev:1994rc} are listed in the table as well.}\label{radiative decay3}
\begin{tabular}{ccccccccccccc}  \midrule[1.0pt]\midrule[1.0pt]
 ~~~~~Initial~~~~~  & ~~~~~Final~~~~~& \multicolumn{5}{c} {\underline{~~~~~~~~~~~~~~~~~~~~~~~~~~~~$E_{\gamma}$  (MeV)~~~~~~~~~~~~~~~~~~~~~~~~~~~~}} & \multicolumn{4}{c} {\underline{~~~~~~~~~~~~~~~~~~~~~$\Gamma_{\mathrm{E1}}$  (keV)~~~~~~~~~~~~~~~~~~~~~}} & \multicolumn{2}{c} {\underline{$\Gamma_{\mathrm{EM}}$  (keV)}}  \\
   ~~~~~state~~~~~     & ~~~~~state~~~~~  &~~~~~\cite{Eichten:1994gt}~~~~~&~~~~~\cite{Ebert:2002pp}~~~~~&~~~~~\cite{Godfrey:2004ya}~~~~~& ~~~~~\cite{Kiselev:1994rc}~~~~~&~~~~~ours~~~~~
   &~~~~~\cite{Eichten:1994gt}~~~~~& ~~~~~\cite{Ebert:2002pp}~~~~~&~~~~~ \cite{Godfrey:2004ya}~~~~~
   & ~~~~~\cite{Kiselev:1994rc}~~~~~ &~~~~~ Ours~~~~~  		\\
\midrule[1.0pt]
$1 ^3P_{2}$        & $1 ^3S_{1}$ & 397   &   416   &   416 & 426 & 445 & 112.6  &  122    &   83   &   102.9   &   87      \\
$1P'_1$            &             & 387   &   405   &   399 & 412 & 433 & 0.1    &  13.7   &   11   &   8.1     &   40      \\
$1P_1 $            &             & 382   &   389   &   391 & 400 & 416 & 99.5   &  87.1   &   60   &   77.8    &   70      \\
$1 ^3P_{0}$        &             & 353   &   355   &   358 & 366 & 377 & 79.2   &  75.5   &   55   &   65.3    &   96      \\
$1P'_1$            & $1 ^1S_{0}$ & 455   &   463   &   462 & 476 & 484 & 56.4   &  147    &   80   &   131.1   &   74      \\
$1P_1 $            &             & 450   &   447   &   454 & 464 & 468 & 0      &  18.4   &   13   &   11.6    &   35      \\
\end{tabular}
 \end{table*}
 
\begin{table*}[htb]
\begin{center}
\caption{\label{radiative decay5} Partial widths of the $E1$ dominant radiative transitions for the  $2D$-wave $B_c$ states.}
\scalebox{1.0}{
\begin{tabular}{ccccccccccccccccccc}
\midrule[1.0pt]\midrule[1.0pt]
~~~&	 Initial state	~~~~&	Final state    ~~~~&	$E_{\gamma}$ (MeV)	~~~~~~~~~~~&	$\Gamma_{\mathrm{EM}}$ (keV)
~~~~~~~~~~~&	 Initial state	~~~~&	Final state    ~~~~&	$E_{\gamma}$ (MeV)	~~~~~~~~~~~&	$\Gamma_{\mathrm{EM}}$ (keV)	\\
\midrule[1.0pt]
~~~&	$2 ^3D_{1}$~~~~~~~~&  $1 ^3P_{2}$ ~~~~~~~~&  528        ~~~~~~~~& 8.13   ~~~~~~~~&  $2 ^3D_{3}$ ~~~~~~~~& $1 ^3P_{2}$ ~~~~~~~~&  540        ~~~&	32	 \\
~~~&              ~~~~~~~~&  $1P'$       ~~~~~~~~&  540        ~~~~~~~~& 7.6    ~~~~~~~~&              ~~~~~~~~& $1P'$       ~~~~~~~~&  552        ~~~&	0.54 \\
~~~&              ~~~~~~~~&  $1P $       ~~~~~~~~&  557        ~~~~~~~~& 12.5   ~~~~~~~~&              ~~~~~~~~& $1P $       ~~~~~~~~&  568        ~~~&	1.23 \\
~~~&              ~~~~~~~~&  $1 ^3P_{0}$ ~~~~~~~~&  596        ~~~~~~~~& 41.8   ~~~~~~~~&              ~~~~~~~~& $1 ^3P_{0}$ ~~~~~~~~&  607        ~~~&	2.04 \\
~~~&              ~~~~~~~~&  $2 ^3P_{2}$ ~~~~~~~~&  174        ~~~~~~~~& 0.58   ~~~~~~~~&              ~~~~~~~~& $2 ^3P_{2}$ ~~~~~~~~&  186        ~~~&	54	 \\
~~~&              ~~~~~~~~&  $2P'$       ~~~~~~~~&  184        ~~~~~~~~& 10.15  ~~~~~~~~&              ~~~~~~~~& $2P'$       ~~~~~~~~&  195        ~~~&	0.09 \\
~~~&              ~~~~~~~~&  $2P $       ~~~~~~~~&  199        ~~~~~~~~& 20.88  ~~~~~~~~&              ~~~~~~~~& $2P $       ~~~~~~~~&  211        ~~~&	0.23 \\
\end{tabular}}
\end{center}
\end{table*}

\begin{table*}[htb]
\caption{ Partial widths of the $E1$ dominant radiative transitions for the  $2S$-, $2P$-wave $B_c$ states.
For comparison, the predictions from the relativistic quark model~\cite{Ebert:2002pp}, relativized quark model~\cite{Godfrey:2004ya}, nonrelativistic constituent quark models~\cite{Eichten:1994gt,Kiselev:1994rc} are listed in the table as well.
}\label{radiative decay2}
\begin{tabular}{ccccccccccccc}  \midrule[1.0pt]\midrule[1.0pt]
 ~~~~~Initial~~~~~  & ~~~~~Final~~~~~& \multicolumn{5}{c} {\underline{~~~~~~~~~~~~~~~~~~~~~~~~~~~~$E_{\gamma}$  (MeV)~~~~~~~~~~~~~~~~~~~~~~~~~~~~}} & \multicolumn{4}{c} {\underline{~~~~~~~~~~~~~~~~~~~~~$\Gamma_{\mathrm{E1}}$  (keV)~~~~~~~~~~~~~~~~~~~~~}} & \multicolumn{2}{c} {\underline{$\Gamma_{\mathrm{EM}}$  (keV)}}  \\
   ~~~~~state~~~~~     & ~~~~~state~~~~~  &~~~~~\cite{Eichten:1994gt}~~~~~&~~~~~\cite{Ebert:2002pp}~~~~~&~~~~~\cite{Godfrey:2004ya}~~~~~& ~~~~~\cite{Kiselev:1994rc}~~~~~&~~~~~ours~~~~~
   &~~~~~\cite{Eichten:1994gt}~~~~~& ~~~~~\cite{Ebert:2002pp}~~~~~&~~~~~ \cite{Godfrey:2004ya}~~~~~
   & ~~~~~\cite{Kiselev:1994rc}~~~~~ &~~~~~ Ours~~~~~  										\\
            \midrule[1.0pt]      
$2 ^3S_{1}$        & $1 ^3P_{2}$ & 151   &   118   &   118 & 159 & 102 & 17.7   &  7.59   &   5.7  &   14.8    &   6.98   \\
                   & $1P'_1$     & 161   &   130   &   136 & 173 & 115 &   0    &  0.74   &   0.7  &   1.0     &   1.56    \\
                   & $1P_1$      & 167   &   146   &   144 & 185 & 133 & 14.5   &  7.65   &   4.7  &   12.8    &   4.62    \\
                   & $1 ^3P_{0}$ & 196   &   181   &   179 & 219 & 174 & 7.8    &  5.53   &   2.9  &   7.7     &   3.48   \\
$2 ^1S_{0}$        & $1P'_1$     & 119   &   84    &   104 & 138 & 96  & 5.2    &  4.40   &   6.1  &   15.9    &   6.38    \\
                   & $1P_1 $     & 125   &   101   &   113 & 150 & 114 & 0      &  1.05   &   1.3  &   1.9     &   5.33    \\
\end{tabular}
 \end{table*}

\begin{table*}[htp]
\begin{center}
\caption{\label{radiative decay55} Partial widths of the $E1$ dominant radiative transitions for the $3S$-, $4S$-, $3P$-wave $B_c$ states.}
\scalebox{1.0}{
\begin{tabular}{cccccccccccccccccccccccccccccccc}
\midrule[1.0pt]\midrule[1.0pt]
~~~&	 Initial state	~~~~&	Final state	    ~~~~&	$E_{\gamma}$ (MeV)	~~~~~~~~~~~&	$\Gamma_{\mathrm{EM}}$ (keV)
~~~~~~~~~~~&	 Initial state	~~~~&	Final state	    ~~~~&	$E_{\gamma}$ (MeV)	~~~~~~~~~~~&	$\Gamma_{\mathrm{EM}}$ (keV)	\\
\midrule[1.0pt]
~~~&  $3 ^1S_{0}$ ~~~~~~~~&  $2P'$       ~~~~~~~~&  88         ~~~~~~~~& 11.13  ~~~~~~~~&  $3 ^3S_{1}$ ~~~~~~~~& $2 ^3P_{2}$ ~~~~~~~~&  91         ~~~~~~~~&	11.89\\
~~~&              ~~~~~~~~&  $2P $       ~~~~~~~~&  104        ~~~~~~~~& 10.93  ~~~~~~~~&              ~~~~~~~~& $2P'$       ~~~~~~~~&  101        ~~~~~~~~&	2.92 \\
~~~&              ~~~~~~~~&  $1P'$       ~~~~~~~~&  450        ~~~~~~~~& 1.74   ~~~~~~~~&              ~~~~~~~~& $2P $       ~~~~~~~~&  117        ~~~~~~~~&	7.2  \\
~~~&              ~~~~~~~~&  $1P $       ~~~~~~~~&  467        ~~~~~~~~& 1.25   ~~~~~~~~&              ~~~~~~~~& $2 ^3P_{0}$ ~~~~~~~~&  144        ~~~~~~~~&	5    \\
~~~&              ~~~~~~~~&              ~~~~~~~~&             ~~~~~~~~&        ~~~~~~~~&              ~~~~~~~~& $1 ^3P_{2}$ ~~~~~~~~&  450        ~~~~~~~~&	1.58 \\
~~~&              ~~~~~~~~&              ~~~~~~~~&             ~~~~~~~~&        ~~~~~~~~&              ~~~~~~~~& $1P'$       ~~~~~~~~&  462        ~~~~~~~~&	0.7  \\
~~~&              ~~~~~~~~&              ~~~~~~~~&             ~~~~~~~~&        ~~~~~~~~&              ~~~~~~~~& $1P $       ~~~~~~~~&  479        ~~~~~~~~&	1.72 \\
\end{tabular}}
\end{center}
\end{table*}


\begin{table*}[htp]
\begin{center}
\caption{\label{strong decay5} Strong decay properties for the $4S$-, $5S$-wave $B_c$ states. $\Gamma_{th}$ and $B_{r}$ stand for the partial widths and branching ratios of the strong decay processes, respectively.}
\scalebox{1.0}{
\begin{tabular}{ccccccccccccccc}\midrule[1.0pt]\midrule[1.0pt]
& ~~~State	~~~~&	Decay mode~~~ &	$\Gamma_{th}$ (MeV)~~~~~~~                        &	$B_{r}(\%)$ ~~~~~~~                         	
& ~~~State	~~~&	Decay mode~~~ &	$\Gamma_{th}$ (MeV)~~~~~~~                        &	$B_{r}(\%)$ ~~~~~~~                         	                                                   \\
\midrule[1.0pt]
&~~~~~~~$3 ^1S_{0}$(7239)~~~~~~~& $B^{*}D$                         ~~~~~~~&  161    ~~~~~~~&  100   ~~~~~~~&   ~~~~~~~$3 ^3S_{1}$(7252)~~~~~~~& $BD$                         ~~~~~~~&  28     ~~~~~~~&21     ~~~~~~~&\\
&~~~~~~~                 ~~~~~~~&                                  ~~~~~~~&         ~~~~~~~&        ~~~~~~~&   ~~~~~~~                 ~~~~~~~& $B^{*}D$                     ~~~~~~~&  105    ~~~~~~~&79     ~~~~~~~&\\
&~~~~~~~                 ~~~~~~~& Total                            ~~~~~~~&  161    ~~~~~~~&  100   ~~~~~~~&   ~~~~~~~                 ~~~~~~~& Total                        ~~~~~~~&  133    ~~~~~~~&100    ~~~~~~~&\\ \hline
&~~~~~~~$4 ^1S_{0}$(7540)~~~~~~~& $B^{*}D$                         ~~~~~~~&  0.14   ~~~~~~~&  0.1   ~~~~~~~&   ~~~~~~~$4 ^3S_{1}$(7550)~~~~~~~& $BD$                         ~~~~~~~&  4.53   ~~~~~~~&2.7    ~~~~~~~&\\
&~~~~~~~                 ~~~~~~~& $BD^{*}$                         ~~~~~~~&  34.9   ~~~~~~~&  18.3  ~~~~~~~&   ~~~~~~~                 ~~~~~~~& $B^{*}D$                     ~~~~~~~&  0.41   ~~~~~~~&0.2    ~~~~~~~&\\
&~~~~~~~                 ~~~~~~~& $B^{*}D^{*}$                     ~~~~~~~&  104    ~~~~~~~&  54    ~~~~~~~&   ~~~~~~~                 ~~~~~~~& $BD^{*}$                     ~~~~~~~&  17.0  ~~~~~~~&10     ~~~~~~~&\\
&~~~~~~~                 ~~~~~~~& $B_{s}^{*0}D_{s}^{+}$            ~~~~~~~&  6.7    ~~~~~~~&  3.5   ~~~~~~~&   ~~~~~~~                 ~~~~~~~& $B^{*}D^{*}$                 ~~~~~~~&  112    ~~~~~~~&66     ~~~~~~~&\\
\end{tabular}}
\end{center}
\end{table*}

\begin{table*}[htp]
\begin{center}
\caption{\label{strong decay6} Strong decay properties for the $6S$-wave $B_c$ states.}
\scalebox{1.0}{
\begin{tabular}{ccccccccccccccc}\midrule[1.0pt]\midrule[1.0pt]
& ~~~~~State	~~~~~&	Decay mode~~~~~&	$\Gamma_{th}$ (MeV)~~~~~                        &	$B_{r}(\%)$ ~~~~~                         	
& ~~~~~State	~~~~~&	Decay mode~~~~~&	$\Gamma_{th}$ (MeV)~~~~~                        &	$B_{r}(\%)$ ~~~~~         \\
\midrule[1.0pt]
&~~~~~$6 ^1S_{0}$(8046)~~~~~& $B^{*}D$                    ~~~~~&  44.4  ~~~~~&12    ~~~~~& ~~~~~$6 ^3S_{1}$(8054)~~~~~& $BD$                        ~~~~~&  17.6  ~~~~~&4.7   \\
&~~~~~                 ~~~~~& $BD^{*}$                    ~~~~~&  24.3  ~~~~~&6.7   ~~~~~& ~~~~~                 ~~~~~& $B^{*}D$                    ~~~~~&  31     ~~~~~&8.3   \\
&~~~~~                 ~~~~~& $B^{*}D^{*}$                ~~~~~&  24.3  ~~~~~&6.7   ~~~~~& ~~~~~                 ~~~~~& $BD^{*}$                    ~~~~~&  19.1   ~~~~~&5.1   \\
&~~~~~                 ~~~~~& $B_{s}^{*0}D_{s}^{+}$       ~~~~~&  1.11   ~~~~~&0.3   ~~~~~& ~~~~~                 ~~~~~& $B^{*}D^{*}$                ~~~~~&  37.9   ~~~~~&10.2  \\
&~~~~~                 ~~~~~& $B_{s}^{0}D_{s}^{*+}$       ~~~~~&  0.38   ~~~~~&0.11  ~~~~~& ~~~~~                 ~~~~~& $B_{s}^{0}D_{s}^{+}$        ~~~~~&  1.78   ~~~~~&0.5   \\
&~~~~~                 ~~~~~& $B_{s}^{*0}D_{s}^{*+}$      ~~~~~&  3.33   ~~~~~&0.9   ~~~~~& ~~~~~                 ~~~~~& $B_{s}^{*0}D_{s}^{+}$       ~~~~~&  1.22   ~~~~~&0.3   \\
&~~~~~                 ~~~~~& $B(1^{3}P_{0})D$            ~~~~~&  11.3  ~~~~~&3.1   ~~~~~& ~~~~~                 ~~~~~& $B_{s}^{0}D_{s}^{*+}$       ~~~~~&  0.09   ~~~~~&0.02  \\
\end{tabular}}
\end{center}
\end{table*}

\begin{table*}[htp]
\begin{center}
\caption{\label{strong decay7} Strong decay properties for the $3P$-, $4P$-wave $B_c$ states.}
\scalebox{1.0}{
\begin{tabular}{ccccccccccccccc}\midrule[1.0pt]\midrule[1.0pt]
& ~~~~~~~State	~~~~~~~&	Decay mode~~~~~~~&	$\Gamma_{th}$ (MeV)~~~~~~~                        &	$B_{r}(\%)$ ~~~~~~~                         	
& ~~~~~~~State	~~~~~~~&	Decay mode~~~~~~~&	$\Gamma_{th}$ (MeV)~~~~~~~                        &	$B_{r}(\%)$ ~~~~~~~                         	                                                   \\
\midrule[1.0pt]
  &~~~~~~~$3 ^3P_{0}$(7420)~~~~~~~& $BD$                              ~~~~~~~&  9.6    ~~~~~~~&3.5    ~~~~~~~& ~~~~~~~$3 ^3P_{2}$(7464)~~~~~~~& $BD$                         ~~~~~~~&  22     ~~~~~~~&11.1   ~~~~~~~&\\
  &~~~~~~~                 ~~~~~~~& $B^{*}D^{*}$                      ~~~~~~~&  255    ~~~~~~~&93     ~~~~~~~& ~~~~~~~                 ~~~~~~~& $B^{*}D$                     ~~~~~~~&  16     ~~~~~~~&8.1    ~~~~~~~&\\
  &~~~~~~~                 ~~~~~~~& $B_{s}^{0}D_{s}^{+}$              ~~~~~~~&  9.7   ~~~~~~~&3.5    ~~~~~~~& ~~~~~~~                 ~~~~~~~& $BD^{*}$                     ~~~~~~~&  3.4   ~~~~~~~&1.7    ~~~~~~~&\\
  &~~~~~~~                 ~~~~~~~&                                   ~~~~~~~&         ~~~~~~~&       ~~~~~~~& ~~~~~~~                 ~~~~~~~& $B^{*}D^{*}$                 ~~~~~~~&  146    ~~~~~~~&74     ~~~~~~~&\\
  &~~~~~~~                 ~~~~~~~&                                   ~~~~~~~&         ~~~~~~~&       ~~~~~~~& ~~~~~~~                 ~~~~~~~& $B_{s}^{0}D_{s}^{+}$         ~~~~~~~&  2.7    ~~~~~~~&1.4    ~~~~~~~&\\
  &~~~~~~~                 ~~~~~~~&                                   ~~~~~~~&         ~~~~~~~&       ~~~~~~~& ~~~~~~~                 ~~~~~~~& $B_{s}^{*0}D_{s}^{+}$        ~~~~~~~&  7.8    ~~~~~~~&4      ~~~~~~~&\\
  &~~~~~~~                 ~~~~~~~& Total                             ~~~~~~~&  274 ~~~~~~~&100    ~~~~~~~& ~~~~~~~                 ~~~~~~~& Total                        ~~~~~~~&  198 ~~~~~~~&100    ~~~~~~~&\\
\end{tabular}}
\end{center}
\end{table*}

\begin{table*}[htp]
\begin{center}
\caption{\label{strong decay8} Strong decay properties for the $2D$-, $3D$-wave $B_c$ states.}
\scalebox{1.0}{
\begin{tabular}{ccccccccccccccc}\midrule[1.0pt]\midrule[1.0pt]
& ~~~~~~~State	~~~~~~~&	Decay mode~~~~~~~&	$\Gamma_{th}$ (MeV)~~~~~~~                        &	$B_{r}(\%)$ ~~~~~~~                         	
& ~~~~~~~State	~~~~~~~&	Decay mode~~~~~~~&	$\Gamma_{th}$ (MeV)~~~~~~~                        &	$B_{r}(\%)$ ~~~~~~~                         	                                                   \\
\midrule[1.0pt]
&~~~~~~~$2 ^3D_{1}$(7336)~~~~~~~& $BD$                    ~~~~~~~&  0.55   ~~~~~~~&1.0    ~~~~~~~& ~~~~~~~$2 ^3D_{3}$(7348)~~~~~~~& $BD$                         ~~~~~~~&  41.6  ~~~~~~~& 22.1  ~~~~~~~&\\
&~~~~~~~                 ~~~~~~~& $B^{*}D$                ~~~~~~~&  6.24   ~~~~~~~&10.9   ~~~~~~~& ~~~~~~~                 ~~~~~~~& $B^{*}D$                     ~~~~~~~&  50.8  ~~~~~~~& 26.9  ~~~~~~~&\\
&~~~~~~~                 ~~~~~~~& $BD^{*}$                ~~~~~~~&  50.1  ~~~~~~~&87     ~~~~~~~& ~~~~~~~                 ~~~~~~~& $BD^{*}$                     ~~~~~~~&  9.29   ~~~~~~~& 4.9   ~~~~~~~&\\
&~~~~~~~                 ~~~~~~~& $B^{*}D^{*}$            ~~~~~~~&  0.48   ~~~~~~~&0.8    ~~~~~~~& ~~~~~~~                 ~~~~~~~& $B^{*}D^{*}$                 ~~~~~~~&  87     ~~~~~~~& 46.1  ~~~~~~~&\\
&~~~~~~~                 ~~~~~~~& $B_{s}^{0}D_{s}^{+}$    ~~~~~~~&  0.18   ~~~~~~~&0.3    ~~~~~~~& ~~~~~~~                 ~~~~~~~& $B_{s}^{0}D_{s}^{+}$         ~~~~~~~&  0.013  ~~~~~~~& 0.01  ~~~~~~~&\\
&~~~~~~~                 ~~~~~~~& Total                   ~~~~~~~&  57   ~~~~~~~&100    ~~~~~~~& ~~~~~~~                 ~~~~~~~& Total                        ~~~~~~~&  189 ~~~~~~~& 100   ~~~~~~~&\\ \hline
&~~~~~~~$2D_{2}'$(7347)  ~~~~~~~& $B^{*}D$                ~~~~~~~&  57.1  ~~~~~~~&34.7   ~~~~~~~& ~~~~~~~$2 D_{2}$(7343)  ~~~~~~~& $B^{*}D$                     ~~~~~~~&  38.2  ~~~~~~~&27     ~~~~~~~&\\
\end{tabular}}
\end{center}
\end{table*}

\begin{table*}[htp]
\begin{center}
\caption{\label{strong decay9} Strong decay properties for the $1F$-, $2F$-wave $B_c$ states.}
\scalebox{1.0}{
\begin{tabular}{ccccccccccccccc}\midrule[1.0pt]\midrule[1.0pt]
& ~~~~~~~State	~~~~~~~&	Decay mode~~~~~~~&	$\Gamma_{th}$ (MeV)~~~~~~~                        &	$B_{r}(\%)$ ~~~~~~~                         	
& ~~~~~~~State	~~~~~~~&	Decay mode~~~~~~~&	$\Gamma_{th}$ (MeV)~~~~~~~                        &	$B_{r}(\%)$ ~~~~~~~                         	                                                   \\
\midrule[1.0pt]
&~~~~~~~$1 ^3F_{2}$(7235)~~~~~~~& $BD$                    ~~~~~~~&  61.9  ~~~~~~~&85      ~~~~~~~& ~~~~~~~$1 ^3F_{4}$(7227)~~~~~~~& $BD$                         ~~~~~~~&  0.85   ~~~~~~~& 97    ~~~~~~~&\\
&~~~~~~~                 ~~~~~~~& $B^{*}D$                ~~~~~~~&  11.1   ~~~~~~~&15      ~~~~~~~& ~~~~~~~                 ~~~~~~~& $B^{*}D$                     ~~~~~~~&  0.03   ~~~~~~~& 3     ~~~~~~~&\\
&~~~~~~~                 ~~~~~~~& Total                   ~~~~~~~&  73   ~~~~~~~&100     ~~~~~~~& ~~~~~~~                 ~~~~~~~& Total                        ~~~~~~~&  0.88   ~~~~~~~& 100   ~~~~~~~&\\ \hline
&~~~~~~~$1F_{3}'$(7240)  ~~~~~~~& $B^{*}D$                ~~~~~~~&  15.1  ~~~~~~~&  100   ~~~~~~~& ~~~~~~~$1 F_{3}$(7224)  ~~~~~~~& $B^{*}D$                     ~~~~~~~&  8.53   ~~~~~~~&  100  ~~~~~~~&\\
&~~~~~~~                 ~~~~~~~& Total                   ~~~~~~~&  15.1  ~~~~~~~&  100   ~~~~~~~& ~~~~~~~                 ~~~~~~~& Total                        ~~~~~~~&  8.53   ~~~~~~~&  100  ~~~~~~~&\\ \hline
&~~~~~~~$2 ^3F_{2}$(7518)~~~~~~~& $BD$                    ~~~~~~~&  45.1  ~~~~~~~&20.2    ~~~~~~~& ~~~~~~~$2 ^3F_{4}$(7514)~~~~~~~& $BD$                         ~~~~~~~&  8      ~~~~~~~& 6     ~~~~~~~&\\
&~~~~~~~                 ~~~~~~~& $B^{*}D$                ~~~~~~~&  19.2  ~~~~~~~&8.6     ~~~~~~~& ~~~~~~~                 ~~~~~~~& $B^{*}D$                     ~~~~~~~&  20.9  ~~~~~~~& 16    ~~~~~~~&\\
\end{tabular}}
\end{center}
\end{table*}

\begin{table*}[htp]
\begin{center}
\caption{\label{strong decay10} Strong decay properties for the $3F$-wave $B_c$ states.}
\scalebox{1.0}{
\begin{tabular}{ccccccccccccccc}\midrule[1.0pt]\midrule[1.0pt]
& ~~~~~~~State	~~~~~~~&	Decay mode~~~~~~~&	$\Gamma_{th}$ (MeV)~~~~~~~                        &	$B_{r}(\%)$ ~~~~~~~                         	
& ~~~~~~~State	~~~~~~~&	Decay mode~~~~~~~&	$\Gamma_{th}$ (MeV)~~~~~~~                        &	$B_{r}(\%)$ ~~~~~~~                         	                                                   \\
\midrule[1.0pt]
&~~~~~~~$3 ^3F_{2}$(7730)~~~~~~~& $BD$                        ~~~~~~~&  32.1  ~~~~~~~& 14     ~~~~~~~& ~~~~~~~$3 ^3F_{4}$(7771)~~~~~~~& $BD$                         ~~~~~~~&  2.82   ~~~~~~~& 1.6    ~~~~~~~&\\
&~~~~~~~                 ~~~~~~~& $B^{*}D$                    ~~~~~~~&  16.1  ~~~~~~~& 7      ~~~~~~~& ~~~~~~~                 ~~~~~~~& $B^{*}D$                     ~~~~~~~&  8.9    ~~~~~~~& 5.0    ~~~~~~~&\\
&~~~~~~~                 ~~~~~~~& $BD^{*}$                    ~~~~~~~&  3.89   ~~~~~~~& 1.7    ~~~~~~~& ~~~~~~~                 ~~~~~~~& $BD^{*}$                     ~~~~~~~&  20.2  ~~~~~~~& 11.4   ~~~~~~~&\\
&~~~~~~~                 ~~~~~~~& $B^{*}D^{*}$                ~~~~~~~&  72     ~~~~~~~& 31.5   ~~~~~~~& ~~~~~~~                 ~~~~~~~& $B^{*}D^{*}$                 ~~~~~~~&  50.9  ~~~~~~~& 28.7   ~~~~~~~&\\
&~~~~~~~                 ~~~~~~~& $B_{s}^{0}D_{s}^{+}$        ~~~~~~~&  0.38   ~~~~~~~& 0.2    ~~~~~~~& ~~~~~~~                 ~~~~~~~& $B_{s}^{0}D_{s}^{+}$         ~~~~~~~&  3.2    ~~~~~~~& 1.8    ~~~~~~~&\\
&~~~~~~~                 ~~~~~~~& $B_{s}^{*0}D_{s}^{+}$       ~~~~~~~&  0.11   ~~~~~~~& 0.05   ~~~~~~~& ~~~~~~~                 ~~~~~~~& $B_{s}^{*0}D_{s}^{+}$        ~~~~~~~&  3.2    ~~~~~~~& 1.8    ~~~~~~~&\\
&~~~~~~~                 ~~~~~~~& $B_{s}^{0}D_{s}^{*+}$       ~~~~~~~&  2.09   ~~~~~~~& 0.9    ~~~~~~~& ~~~~~~~                 ~~~~~~~& $B_{s}^{0}D_{s}^{*+}$        ~~~~~~~&  0.4    ~~~~~~~& 0.2    ~~~~~~~&\\
\end{tabular}}
\end{center}
\end{table*}

\section*{  Acknowledgements }

This work is supported by the National Natural Science Foundation of China under Grants No.~11775078, No.~U1832173, No.~11705056, and No.~11405053.

\end{document}